%
%
%

\input harvmac\skip0=\baselineskip

\def\dx{{\dot x}}
\def\s{{\sigma}}
\def\t{{\tau}}
\def\l{{\lambda}}
\def\a{{\alpha}}

\def\p{{\partial}}

\def\pb{{\vec p}}
\def\xb{{\vec x}}
\def\yb{{\vec y}}
\def\msurr{\mathsurround=0pt}
\def\overleftrightarrow#1{\vbox{\msurr\ialign{##\crcr
        $\leftrightarrow$\crcr\noalign{\kern-1pt\nointerlineskip}
        $\hfil\displaystyle{#1}\hfil$\crcr}}}
\noblackbox

\lref\plctmp{ J.~Polchinski, ``Evaluation Of The One Loop String
Path Integral,'' Commun.\ Math.\ Phys.\  {\bf 104}, 37 (1986).}
 \lref\SenAN{ A.~Sen, ``Field theory of tachyon
matter,'' Mod.\ Phys.\ Lett.\ A {\bf 17}, 1797 (2002)
[arXiv:hep-th/0204143]. }

\lref\mpup{S. Minwalla and K. Pappododimas, unpublished.}
\lref\GaberdielXM{ M.~R.~Gaberdiel, A.~Recknagel and G.~M.~Watts,
``The conformal boundary states for SU(2) at level 1,'' Nucl.\
Phys.\ B {\bf 626}, 344 (2002) [arXiv:hep-th/0108102].
} \lref\pdv{ P.~Di Vecchia, M.~Frau, I.~Pesando, S.~Sciuto,
A.~Lerda and R.~Russo, ``Classical p-branes from boundary state,''
Nucl.\ Phys.\ B {\bf 507}, 259 (1997) [arXiv:hep-th/9707068].
} \lref\llm{N. Lambert, H. Liu and J. Maldacena to appear.}
\lref\PolchinskiRQ{ J.~Polchinski, ``String Theory. Vol. 1: An
Introduction To The Bosonic String.'' }

\lref\okd{ T.~Okuda and
S.~Sugimoto, ``Coupling of rolling tachyon to closed strings,''
Nucl.\ Phys.\ B {\bf 647}, 101 (2002) [arXiv:hep-th/0208196].}

\lref\PolchinskiMY{
J.~Polchinski and L.~Thorlacius,
``Free Fermion Representation Of A Boundary Conformal Field Theory,''
Phys.\ Rev.\ D {\bf 50}, 622 (1994)
[arXiv:hep-th/9404008].
}

\lref\GaberdielZQ{
M.~R.~Gaberdiel and A.~Recknagel,
``Conformal boundary states for free bosons and fermions,''
JHEP {\bf 0111}, 016 (2001)
[arXiv:hep-th/0108238].
}
\lref\GubserIU{
S.~S.~Gubser and A.~Hashimoto,
``Exact absorption probabilities for the D3-brane,''
Commun.\ Math.\ Phys.\  {\bf 203}, 325 (1999)
[arXiv:hep-th/9805140].
}

\lref\KlebanovKC{
I.~R.~Klebanov,
``World-volume approach to absorption by non-dilatonic branes,''
Nucl.\ Phys.\ B {\bf 496}, 231 (1997)
[arXiv:hep-th/9702076].
}

\lref\gradsh{
I. S. Gradshteyn and I. M. Ryzhik,
{\it Table of Integrals, Series and Products},
Academic Press (1994).
}

\lref\mclach{
N.~W.~McLachlan, {\it Theory and Application of Mathieu
Functions}, Clarendon Press, Oxford, UK (1947).
}

\lref\GreenGA{
M.~B.~Green and M.~Gutperle,
``Symmetry Breaking at enhanced Symmetry Points,''
Nucl.\ Phys.\ B {\bf 460}, 77 (1996)
[arXiv:hep-th/9509171].}

\lref\mgas{
M.~Gutperle and A.~Strominger,
``Spacelike branes,''
JHEP {\bf 0204}, 018 (2002)
[arXiv:hep-th/0202210].
}
\lref\CallanUB{
C.~G.~Callan, I.~R.~Klebanov, A.~W.~Ludwig and J.~M.~Maldacena,
``Exact solution of a boundary conformal field theory,''
Nucl.\ Phys.\ B {\bf 422}, 417 (1994)
[arXiv:hep-th/9402113].
}

\lref\SenNU{
A.~Sen,
``Rolling tachyon,''
JHEP {\bf 0204}, 048 (2002)
[arXiv:hep-th/0203211].
}

\lref\SenIN{
A.~Sen,
``Tachyon matter,''
arXiv:hep-th/0203265.
}

\lref\SpradlinPW{
M.~Spradlin, A.~Strominger and A.~Volovich,
``Les Houches lectures on de Sitter space,''
arXiv:hep-th/0110007.
}

\lref\SenFTTM{ A.~Sen, ``Field Theory of Tachyon matter,''
arXiv:hep-th/0204143.
}

\lref\FateevIK{
V.~Fateev, A.~B.~Zamolodchikov and A.~B.~Zamolodchikov,
``Boundary Liouville field theory. I: Boundary state and boundary
two-point function,''
arXiv:hep-th/0001012.
}

\lref\TeschnerMD{
J.~Teschner,
``Remarks on Liouville theory with boundary,''
arXiv:hep-th/0009138.
}

\lref\LarsenWC{
F.~Larsen, A.~Naqvi and S.~Terashima,
``Rolling tachyons and decaying branes,''
arXiv:hep-th/0212248.
}

\lref\SenMG{
A.~Sen,
``Non-BPS states and branes in string theory,''
arXiv:hep-th/9904207.
}

\lref\SenVV{
A.~Sen,
``Time evolution in open string theory,''
JHEP {\bf 0210}, 003 (2002)
[arXiv:hep-th/0207105].
}

\lref\senb{
P.~Mukhopadhyay and A.~Sen,
``Decay of unstable D-branes with electric field,''
JHEP {\bf 0211}, 047 (2002)
[arXiv:hep-th/0208142].}

\lref\sena{A. Sen, unpublished.}

\lref\LuER{
H.~Lu, S.~Mukherji and C.~N.~Pope,
``From p-branes to cosmology,''
Int.\ J.\ Mod.\ Phys.\ A {\bf 14}, 4121 (1999)
[arXiv:hep-th/9612224].
}
\lref\HullVG{
C.~M.~Hull,
``Timelike T-duality, de Sitter space, large N gauge theories and  topological field theory,''
JHEP {\bf 9807}, 021 (1998)
[arXiv:hep-th/9806146].
}
\lref\jk{J. Karczmarek and A. Strominger, in progress.}

\lref\FukudaBV{
T.~Fukuda and K.~Hosomichi,
``Super Liouville theory with boundary,''
Nucl.\ Phys.\ B {\bf 635}, 215 (2002)
[arXiv:hep-th/0202032].}

\lref\StromingerPC{ A.~Strominger, ``Open string creation by
s-branes,'' arXiv:hep-th/0209090.
}
\lref\GutperleXF{
M.~Gutperle and A.~Strominger,
``Timelike boundary Liouville theory,''
arXiv:hep-th/0301038.
}
\lref\ChenYQ{
C.~M.~Chen, D.~V.~Gal'tsov and M.~Gutperle,
``S-brane solutions in supergravity theories,''
Phys.\ Rev.\ D {\bf 66}, 024043 (2002)
[arXiv:hep-th/0204071].}

\lref\MukohyamaCN{
S.~Mukohyama,
``Brane cosmology driven by the rolling tachyon,''
Phys.\ Rev.\ D {\bf 66}, 024009 (2002)
[arXiv:hep-th/0204084].}

\lref\KruczenskiAP{
M.~Kruczenski, R.~C.~Myers and A.~W.~Peet,
``Supergravity S-branes,''
JHEP {\bf 0205}, 039 (2002)
[arXiv:hep-th/0204144].
}

\lref\McInnesNM{
B.~McInnes,
``dS/CFT, censorship, instability of hyperbolic horizons, and spacelike  branes,''
arXiv:hep-th/0205103.
}

\lref\RoyIK{
S.~Roy,
``On supergravity solutions of space-like Dp-branes,''
JHEP {\bf 0208}, 025 (2002)
[arXiv:hep-th/0205198].
}

\lref\DegerIE{
N.~S.~Deger and A.~Kaya,
``Intersecting S-brane solutions of D = 11 supergravity,''
JHEP {\bf 0207}, 038 (2002)
[arXiv:hep-th/0206057].
}

\lref\WangFD{
J.~E.~Wang,
``Spacelike and time dependent branes from DBI,''
JHEP {\bf 0210}, 037 (2002)
[arXiv:hep-th/0207089].
}

\lref\IvashchukGE{
V.~D.~Ivashchuk,
 ``Composite S-brane solutions related to Toda-type systems,''
Class.\ Quant.\ Grav.\  {\bf 20}, 261 (2003)
[arXiv:hep-th/0208101].
}

\lref\QuevedoTM{
F.~Quevedo, G.~Tasinato and I.~Zavala,
``S-branes, negative tension branes and cosmology,''
arXiv:hep-th/0211031.
}

\lref\OhtaUW{
N.~Ohta,
``Intersection rules for S-branes,''
arXiv:hep-th/0301095.
}
\lref\BurgessGG{
C.~P.~Burgess, P.~Martineau, F.~Quevedo, G.~Tasinato and I.~Zavala C.,
``Instabilities and particle production in S-brane geometries,''
arXiv:hep-th/0301122.
}


\Title{\vbox{\baselineskip12pt\hbox{hep-th/0302146}
\hbox{HUTP-03/A016}
}}{S-Brane Thermodynamics}

\centerline{Alexander Maloney, Andrew
Strominger and Xi Yin}
\smallskip
\centerline{
Jefferson Physical Laboratory, Harvard University, Cambridge, MA }
\vskip .6in \centerline{\bf Abstract} {The description of string-theoretic
s-branes at $g_s=0$ as exact worldsheet CFTs with
a $\lambda \cosh X^0$ or $\lambda e^{\pm X^0}$ boundary interaction
is considered.
Due to the imaginary-time periodicity of the interaction under
$X^0 \to X^0+2\pi i$, these configurations have intriguing similarities to
black hole or de Sitter geometries. For example,
the open string pair production as seen
by an Unruh detector is thermal at temperature $T={1 \over 4 \pi}$.
It is shown that, despite the rapid time
dependence of the s-brane, there exists an exactly thermal mixed state
of open strings. The corresponding boundary state is constructed for both
the bosonic and superstring cases.
This state defines a long-distance Euclidean
effective field theory whose light modes are confined to
the s-brane. At the critical value of the coupling
$\lambda=\half$, the boundary
 interaction  simply generates an $SU(2)$ rotation by $\pi$
from Neumman to
Dirichlet boundary conditions.  The $\lambda=\half$ s-brane reduces to an array of
sD-branes (D-branes with a transverse time dimension) on the imaginary
time axis. The long range force between a (bosonic) sD-brane and an ordinary
D-brane is shown from the annulus diagram to be $11 \over 12$
times the force between two D-branes.
The linearized time-dependent RR field $F_{p+2}=dC_{p+1}$
produced by an sDp-brane in superstring theory is explicitly
computed and found to carry a half unit of s-charge
$Q_s=\int_{\Sigma_{8-p}} *F_{p+2}=\half$, where $\Sigma_{8-p}$ is any
transverse spacelike slice.
} \vskip .3in

\smallskip
\Date{}

\listtoc
\writetoc

\

\vfill \eject

\newsec{Introduction }

  A spacelike brane, or s-brane, is much like an ordinary brane
except that one of its transverse dimensions includes time.
S-branes arise as time-dependent, soliton-like configurations in a
variety of field theories. In string theory, the potential for the
open string tachyon field leads to s-branes \mgas\ in a
time-dependent version of the construction \SenMG\ of D-branes as
solitons of the open string tachyon. These s-branes can be thought
of as the creation and subsequent decay of an unstable brane. They
are of interest as relatively simple examples of time-dependent
string backgrounds. Some of the recent investigations can be found
in \refs{\SenNU \SenIN \ChenYQ \MukohyamaCN \KruczenskiAP
\McInnesNM \RoyIK \DegerIE \WangFD \StromingerPC \IvashchukGE \okd
\QuevedoTM \OhtaUW \GutperleXF -\BurgessGG} and related earlier
work is in \refs{\LuER,\HullVG}. 

An elegant worldsheet
construction of a family of s-branes was given in the classical
$g_s=0$ limit by Sen \SenNU. This construction employs an analytic
continuation of the conformally-invariant boundary Sine-Gordon
model. It describes the tachyon field on an unstable bosonic
D-brane by the boundary interaction on the string worldsheet
\eqn\wxz{S_{boundary}=\lambda \int d\tau \cosh {X^0(\tau)\over
\sqrt {\alpha '}},} where $X^0$ is the time coordinate.
Qualitatively similar constructions were also given for the
superstring \SenIN. In this paper we will describe some surprising
and intriguing properties of these s-branes.

A salient feature of time-dependent backgrounds is that there is
in general no preferred vacuum and particle production is
unavoidable. We study the open string vacua on an s-brane and find
that they have somewhat mysterious thermal properties reminiscent
of black hole or de Sitter vacua. For the case \wxz\ there is open
string pair production with a strength characterized by the
Hagedorn temperature \refs{\StromingerPC, \GutperleXF}
\eqn\asd{T_H={1 \over 4\pi \sqrt {\alpha '}}.}
Mathematically, the temperature
arises from the periodicity of the boundary interaction \wxz\ in
imaginary time. Physically, we show that (for the quantum state of
a brane created with no incoming string excitations) an Unruh
detector will see a thermal bath. We further show that at late
times the exact Green function approaches the thermal Green
function plus asymptotically vanishing corrections.

The appearance of the Hagedorn temperature signals a breakdown of
string perturbation theory \refs{\StromingerPC,\okd ,\llm,\mpup}.
Describing the mechanism which cuts off this divergence is an
interesting problem which we will not address in this paper.\foot{
The divergence might be controlled by lowering the s-brane temperature
either with the addition of an electric field \refs{\sena,\senb}
or a null linear dilaton \jk.} Herein we simply avoid the problem
by working at $g_s=0$. Interestingly, we find that the Hagedorn
problem disappears in the $\lambda=\half$ sD-brane case discussed
below.\foot{This can be seen both from the absence of on-shell
open strings and the vanishing of the projection of the boundary
state onto the massless winding state in the closed string channel
associated to the Hagedorn divergence.}

In addition to the pure vacuum states mentioned above,
we construct a series of
mixed thermal states with temperatures \eqn\asd{
T={1 \over 2\pi n\sqrt {\alpha '}}}
for positive integer $n$. Ordinarily it makes no sense to
discuss a thermal state in a highly time-dependent background.
However, because of the Euclidean time periodicity of the
interaction \wxz, we can construct mixed states whose Green  functions
have exact thermal periodicity at all times. Physically, it is
natural to expect open strings on an
s-brane to be in a mixed state,
since they are correlated with the closed string modes
whose energy was needed to create the s-brane in the first place.

Ordinary branes are usefully characterized at long distances by a
long-distance effective field theory. It is rather subtle to
define such an effective theory for an s-brane because there is no
time-translation invariant ground state around which to define and
expand the low-lying excitations. One may try to define a
long-distance effective field theory as the Euclidean theory which
reproduces the long-distance equal time correlators on the
s-brane. In general these correlators depend strongly on the
quantum state of the fields on the s-brane, and will not behave
like those of any Euclidean field theory whose dimension is that of
the s-brane. However, we show that the thermal state of the
s-brane gives an effective field theory which is essentially a
(twisted) compactification of the unstable brane whose
creation/decay comprises the s-brane. Ultimately, this Euclidean
effective field theory may play an interesting role in timelike holographic
duality.

The thermal s-brane states can be succinctly characterized by a
CFT boundary state.  The thermal boundary state at temperature $T={1 \over 2\pi n\sqrt
{\alpha '}}$ differs from the
zero-temperature boundary state of \SenNU\ by a periodic identification
$\phi \sim \phi +2\pi n$ of
the Euclidean timelike scalar $\phi=-iX^0$. The thermal s-brane
boundary state contains closed strings with winding in the $\phi$
direction. These thermal boundary states enable efficient
evaluation of various worldsheet string diagrams.

At the special value of the coupling $\lambda=\half$ appearing in
\wxz, a dramatic simplification occurs. For any $\lambda$
the boundary interaction
\wxz\ generates a right-moving rotation by $2\pi\lambda$ in the SU(2) level one
current
algebra generated by $\p X^0$, $\cosh X^0$ and $i\sinh X^0$.  When
$\lambda=\half$ the rotation is by
$\pi$, which simply transforms the Neumann boundary state into a
Dirichlet boundary state. The s-brane then degenerates into a
periodic array of sD-branes (i.e. D-branes with a Dirichlet
boundary condition on the time coordinate) located on the
imaginary time axis at $\phi= m \pi $ for odd integer $m$.
This gives the precise relation between the Dirichlet type  boundary states
discussed in \mgas\ and the Sine-Gordon type boundary states discussed in
\SenNU.\foot{Reference \mgas\ only considered sD-branes on the real time axis.
As discussed
therein, these have problems both with the dominant energy condition
and the existence of a well-defined string 
perturbation expansion. Both of these
problems seem to be resolved by moving the sD-brane to imaginary time.
In \okd\ it was shown that, for the bosonic string, the theory with
$\lambda=-\half$ has an sD-brane at real time.}  In this $\lambda=\half$
limit there are no on-shell open string states\foot{Off-shell open string states
still appear in the Euclidean effective field theory.}, but the on-shell closed string states determined from the
boundary state remain. This remaining closed string configuration has the
unusual property that its total energy is of order $1 \over g_s$ and its
behavior can be determined from open string calculations
on
the sD-brane. The annulus diagram connecting one
(bosonic) D-brane and one s-brane is computed for arbitrary
$\lambda$ using the boundary state. At $\lambda=\half$, the
long-distance force between a D-brane and an sD-brane is shown to
be $11 \over 12$ times the force between two D-branes (which
corresponds to $\lambda=0$).\foot{This might appear to contradict the
claim \SenNU\ that $\lambda=\half$ is the trivial closed string vacuum.
The interesting resolution of this apparent conflict,
discussed in section 6.1 in some
detail, is that when there is time-dependence, the boundary state
does not uniquely determine the closed string fields. The
implicit prescription adopted in \SenNU\ differs from the one used
herein, which latter amounts to the use of the Feynman propagators obtained
by analytic continuation from Euclidean space.}

This paper is organized as follows. Section 2 describes various s-brane
vacua and their properties in a minisuperspace approximation
which treats the open strings as quantum fields with a time-dependent
$\cosh t$ or $e^{\pm t}$ mass. Most of the important behavior follows from the
Euclidean periodicity of the interaction, which
is an exact property of the
worldsheet CFT. In section 2.1 we review the ``half s-brane'' corresponding
to brane decay, which is described by a (in some respects simpler)
boundary interaction
$\int e^{X^0}d\tau$ instead of \wxz.
We recall  that open string pair creation is characterized
by the Hagedorn
temperature $T_H={1 \over 4 \pi  \sqrt{\a'}}$ \StromingerPC.
In section 2.2 we consider the time-reverse
process of brane creation, for which a natural state is one with no
incoming open strings.  In section 2.3 we move on to the
full s-brane \wxz, whose linearized solutions are Mathieu functions.
We use these solutions
in section 2.4 to describe two vacua
of the full s-brane with no particles in the far past and the
far future, respectively.
In 2.5 a time-reversal invariant
s-brane vacuum is defined by the condition that there is no particle flux
in the middle of the s-brane at $t=0$.  The resulting positive frequency modes
are found to be bounded in the lower half plane $t\to -i \infty$. This
vacuum can therefore be obtained by analytic continuation from Euclidean
space. Other possible vacua are described in 2.6.

In section 3 we discuss the thermal properties of s-branes. In 3.1
it is shown that thermal Green functions at temperature $T={1
\over 2 \pi n \sqrt{\a'}}$ can be obtained by analytic
continuation from a periodically identified Euclidean section. The
Minkowskian mixed thermal state (or density matrix) which
reproduces these Green functions is explicitly constructed.
Thermal properties of certain pure vacuum states arising from the
imaginary-time periodicity of \wxz\ are also demonstrated. For
example, in 3.2 it is shown that an Unruh detector in  the
vacuum with no incoming particles measures a temperature $T_H$
during brane creation.
Furthermore, in 3.3 it is found that at late times the correlators
in the pure state approach thermal correlators plus asymptotically
vanishing corrections. This suggests that branes are naturally
produced in something close to a thermal state.

Section 4 defines the notion of long-distance effective field theory
for an s-brane. This effective theory is determined as the
Euclideanization and time compactification, with twisted fermion boundary
conditions and a periodic boundary tachyon interaction on the worldsheet,
of the theory on the
unstable brane whose creation/decay describes the s-brane. Long-distance
modes are related to zero modes of this compactified theory.

In section 5.1 the boundary states which generate the finite-temperature
string correlators on an s-brane are constructed. They differ from the
usual expression by a periodic identification of Euclidean time, which
allows for winding modes in the closed string channel. The superstring is
discussed in 5.2, and the allowed temperatures are shown to be
\eqn\asd{T={1 \over 2 \pi n  \sqrt{2\a'}}.}

In section 6 we turn to the special case of $\lambda=\half$, in
which the boundary interaction becomes an $SU(2)$ rotation by
$\pi$.  In this case the s-brane collapses to an array of
sD-branes on the imaginary time axis at $t=m\pi i $ for odd
integer m. In section 6.1 we explain an ambiguity in the
propagator used to obtain the spacetime closed string fields from
the boundary state, which is due to the appearance of on-shell
states in the boundary state. It is shown that the analytic
continuation of the Feynman propagator gives a non-zero answer
even when the support of the boundary state moves off the real
time axis. The summation over the sD-branes $m$ is performed in a
simple example to give a closed-form expression for a massless
closed string field. In 6.2 we compute the RR field emanating from
an sD-brane and integrate it over a transverse spacelike surface
to find the s-charge. Long range dilaton/graviton fields are computed in
6.3. In section 6.4 we compute the annulus diagram connecting an
ordinary D-brane to an s-brane for general $\lambda$, using 
old results on the boundary
Sine-Gordon model \CallanUB . We find that the long distance
force between a D-brane and an s-brane 
has a coefficient of $23+\cos (2\pi \lambda)$,
indicating that the force between an sDp-brane ($\lambda=\half$) and a
D-brane is $11 \over 12$
times the force between two ordinary D-branes. In 6.5 we extend the annulus
computation to finite temperature. In 6.6 we discuss the relation
between sD-branes and D-instantons. 
Finally, we conclude in section
7 with speculations on timelike holography.

Results overlapping with those of this paper will appear in \llm.

\newsec{Quantum vacuum states}

We wish to understand the dynamics of the open string worldsheet
theory with a time-dependent tachyon
 \eqn\sdef{ S=-{1\over
4\pi}\int_{\Sigma_2} d^2\sigma \p^a X^\mu \p_a X_\mu + \int_{\p
\Sigma_2} d\tau \ m^2(X^0) , } where here and henceforth we set  $\a'=1$. 
For the open bosonic string $m^2=T$
where $T$ is the spacetime tachyon, while for the open superstring
$m^2 \sim T^2$ after integrating out worldsheet fermions. We use
the symbol $m^2$ to denote the interaction because the coupling
(among other effects) imparts a mass to the open string states. We
consider three interesting cases described by the marginal
interactions \refs{\SenNU,\StromingerPC,\GutperleXF,\LarsenWC}\foot{For the
bosonic string one could also consider a $\sinh X^0$ interaction,
although we do not do so here. In the bosonic theory $\cosh X^0$
describes a process in which a tachyon rolls up the barrier and
then  back down the same side, while  $\sinh X^0$ describes
the tachyon rolling over the barrier. The bosonic $\sinh X^0$ interaction is
challenging because it ventures into the unbounded side of the
tachyon potential. The superstring potential is always positive,
so there is no analog of a $\sinh X^0$ type interaction.}
\eqn\tpp{ m^2_+(X^0) = {\lambda \over 2} e^{X^0}} \eqn\tmm{
m^2_-(X^0) = {\lambda \over 2} e^{- {X^0}} } \eqn\tis{ m^2_s(X^0)
= \l \cosh X^0  .}
The first case $m_+^2$ describes the
process of brane decay, in which an unstable brane decays via
tachyon condensation. The second case describes the time-reverse
process of brane creation, in which an unstable brane emerges from
the vacuum. The final case describes an s-brane, which is
the process of brane creation followed by brane decay. Brane decay
\tpp\ (creation \tmm) can be thought of as the future (past) half of an
s-brane, i.e. as the limiting
case where the middle of the s-brane is pushed into the infinite past
(future).

An exact CFT analysis of these s-brane theories
should be possible \refs{\SenNU,\StromingerPC,\GutperleXF,\LarsenWC},
and some exact results are given in sections 5 and 6. However in this
and the next section we shall confine ourselves to the minisuperspace
analysis \StromingerPC\ in which the effect of the interaction is simply to
give a time-dependent shift (given by \tpp-\tis ) to the masses of
all the open string states. The range of validity of the minisuperspace
 approximation is
unclear, although some evidence in favor of its validity at high
frequencies was found in \GutperleXF.  However, most of the results of
the next two sections follow from the periodicity in imaginary time, which is
an exact property of the CFTs defined by \tpp-\tis , so we expect our
conclusions to be qualitatively correct.

In the minisuperspace approximation
only the zero-mode dependence of the interaction $m^2(X^0)$ is
considered.  In this case we can plug in the usual mode
solution for the free open string with oscillator number $N$ to
get an effective action for the zero modes \eqn\slow{ S = \int
d\tau \left[-{1\over 4 }\dx^\mu \dx_\mu + {(N-1)} +
2m^2(x^0)\right] .} This is the action of a point particle with a
time dependent mass. Here $x^\mu (\tau)$ is the zero mode part of
$X^{\mu} (\s,\t)$, and the second term in \slow\ is an effective
contribution from the oscillators, including the usual normal
ordering constant. From \slow\ we can write down the Klein-Gordon
equation for the open string wave function $\phi(t,\xb)$,
\eqn\zasd{ \left(\p^\mu\p_\mu - 2{m^2(t)} -({N-1})
\right)\phi(t,\xb)=0, } where $(t, \vec x)$ are the spacetime
coordinates corresponding to the worldsheet fields $(X^0,\vec
X)$. This is the equation of motion for a scalar field with
time-dependent mass.

At this point, we should make a few remarks about field theories
with time dependent mass. Time translation invariance has been
broken, so energy is not conserved and there is no preferred set
of positive frequency modes. This is  a familiar circumstance in
the study of quantum field theories in time-dependent backgrounds
which leads to particle creation. The probability current
$j_\mu= i (\phi^*\p_\mu \phi - \p_\mu \phi^* \phi)$ is 
conserved, allowing us to define the Klein-Gordon inner product
\eqn\kgcur{ \langle f|g\rangle = i \int_\Sigma d \Sigma^\mu
(f^*\p_\mu g - \p_\mu f^* g) } where $\Sigma$ is a spacelike
slice.  This norm does not depend on the choice of $\Sigma$ if $f$
and $g$ solve the wave equation. Normalized positive frequency
modes are chosen to have $\langle f|f\rangle=1$. Negative
frequency modes are complex conjugates of positive frequency
modes, with $\langle f^*| f^*\rangle=-1$. There is a set raising
and lowering operators associated to each choice of mode
decomposition -- these operators obey the usual oscillator algebra
if the corresponding modes are normalized with respect to \kgcur.
We also define a vacuum state associated to each mode
decomposition -- it is the state annihilated by the corresponding
lowering operators.

\subsec{The $|in \rangle_+$ vacuum for brane decay}

In this section we review some results of \StromingerPC\ on a
scalar field with mass \eqn\tppp{ m^2_+(t) = {\lambda \over 2}e^t ,} describing open
strings on a decaying brane. A natural vacuum in this case is that
with no particles present in the far past: we shall denote this
state  $|in \rangle_+$.

Expanding $\phi$ in plane waves \eqn\asd{ \phi
(t,\xb)=e^{i\pb\cdot\xb} u(t) } the wave equation becomes
\eqn\halfeq{\eqalign{ &(\p^2_t + \l e^{t} +\omega^2) u=0,~~~~~
\omega^2 =p^2+{N-1} .}} This is a form of Bessel's equation. It
has normalized, positive frequency solutions \foot{ A $+$ ($-$)
subscript on a wave function denotes solutions with $m^2=m_+^2$
($m^2=m_-^2$) during brane decay (creation).
A wave function without a subscript refers to solutions for full
s-brane $m^2 = m^2_s$. A superscript $in$, $out$ and $0$ on a wave
function denotes solutions that are purely positive frequency when
$t\to -\infty$, $t\to+\infty$ or $t=0$. The wave functions $u$
depend on $\pb$ as well, although we will typically suppress
momentum indices. } \eqn\sols{\eqalign{ u^{in}_+ =\l^{i\omega}
{\Gamma(1-2i\omega)\over\sqrt{2\omega}}
J_{-2i\omega}(2\sqrt{{\l}}e^{t/2})& }} These solutions have been
chosen because they approach flat space positive frequency plane
waves in the far past $t\to-\infty$, \eqn\asd{ u^{in}_+ \sim
{1\over \sqrt{2\omega}} e^{-i\omega t}.} We will also consider the
wave functions \eqn\sols{\eqalign{ u_+^{out} &= \sqrt{ \pi\over2}
(i e^{2\pi\omega})^{-1/2} H^{(2)}_{-2i\omega}(2\sqrt{{\l}}e^{t/2})
}} that are purely positive frequency in the far future
$t\to+\infty$, \eqn\asda{ u_+^{out} \sim {{\l}^{-1/4}\over
\sqrt{2}} \exp\left\{-t/4-2i\sqrt{{\l}}e^{t/2}\right\} .} These
are related to the previous set of wave functions by a Bogolubov
transformation \eqn\bogone{\eqalign{ u_+^{out} &= a u_+^{in} +b
u_+^{in}{}^* }} whose coefficients \eqn\candd{\eqalign{
a=e^{2\pi\omega+\pi i/2}b^*= \sqrt{ \omega \pi}
e^{\pi\omega-\pi i/4}\left({{\l}^{-i\omega}\over \sinh2\pi\omega
\Gamma(1-2i\omega)} \right) }} obey the usual unitarity relation
$|a|^2-|b|^2=1$. All solutions of the wave equation during brane
decay vanish exponentially in the far future (but not in the far
past) because the mass is growing exponentially.

The relation \bogone\ between in and out modes implies as usual
the relation between in and out creation and annihilation
operators \eqn\asdb{\eqalign{ a^{in} &= a a^{out}+b^*
(a^{out})^\dagger.}} From this, the condition that $a^{in}
|in\rangle=0$ implies that $|in\rangle$ is a squeezed state
\eqn\squeezed{\eqalign{ |in\rangle_+&= \Pi_\pb
(1-|\gamma|^2)^{1/4} \exp\left\{-{1\over2}\gamma
(a_\pb^{out}{}^\dagger)^2\right\} |out\rangle_+, ~~~~~~ \gamma =
b^*/a .}} Physically, this is the statement that particles are
produced during brane decay: if we start in a state with no
particles at $t\to-\infty$, there will be many particles at time
$t\to+\infty$. We should emphasize here that $\gamma$ is a
function of $\pb$ \eqn\asdc{ \gamma=b^*/a=-ie^{-2\pi\omega} } that
decreases exponentially as the energy $\omega$ increases. In
particular, this implies that the $|in\rangle_+$ and $|out\rangle_+$ vacua
become identical at very short distances.
The density  of particles with momentum $\pb$ is
\eqn\asdd{ n_\pb = |\gamma|^2=e^{-4\pi\omega} .} Despite the fact
that \squeezed\ is a pure state, this is precisely the Boltzmann
density of states at temperature $T_H=1/4\pi$. In string units,
$T_H$  is the Hagedorn temperature. The fact that this
``temperature'' is so high means that $g_s$ corrections are likely
qualitatively important even for $g_s \to 0$ \StromingerPC , but
we do not consider these here.

The appearance of the temperature $T_H$ here is ultimately due to
the Euclidean periodicity of the interactions \tpp-\tis. This will
lead to other thermal properties  as described in the
next section. Since the thermal periodicity is an exact property
of the worldsheet CFT \sdef, we expect this behavior to persist
beyond the minisuperspace approximation considered here.

\subsec{The $|in \rangle_-$ vacuum for brane creation} Solutions
$u_-$ of the wave equation during brane creation are related to
those ($u_+$) during brane decay by time reversal. In particular,
\eqn\asdf{ u_-^{out} (t) = u_+^{in}(-t) {}^* } becomes a plane
wave in the far future $t\to+\infty$, and \eqn\asdg{ u_-^{in} (t)
= u_+^{out} (-t) ^* =\sqrt{ \pi\over2} (i e^{2\pi\omega})^{1/2}
H^{(1)}_{-2i\omega}(2\sqrt{{\l}}e^{-t/2}) } becomes purely
positive frequency in the far past $t\to-\infty$
\eqn\solsa{\eqalign{ u_-^{in} & \sim {{\l}^{-1/4}\over\sqrt{2}}
\exp\left\{t/4+ 2i\sqrt{{\l}}e^{-t/2} \right\} .}} These two
solutions are related by the Bogolubov transformation
\eqn\bogtwo{\eqalign{ u_-^{in} &= a^* u_-^{out} +b^*
u_-^{out}{}^* ,}}
where $a$ and $b$ are given by \candd.
Wave functions during brane creation vanish
exponentially in the far past but not in the far future, because
the masses are infinite in the far past.

The natural vacuum state during brane creation is not, however,
the time reverse $\cal T$ of the in state for brane decay ${\cal
T}|in \rangle_+$. The latter state has particles present in the
far past with infinite masses. Indeed it would cost an infinite
amount of energy to prepare such an initial state. Rather, the
natural in state $|in \rangle_-$ for brane creation has no
particles in the far past and is the time reverse ${\cal T}|out
\rangle_+$ of the out vacuum for brane decay. We can write $|in \rangle_-$
in terms of the free out operators as
\eqn\sqzed{\eqalign{ |in\rangle_-&= \Pi_\pb (1-|\gamma|^2)^{1/4}
\exp\left\{{1\over2}\gamma^* (a_\pb^{out}{}^\dagger)^2\right\}
|out\rangle_- }} where $\gamma$ is given by \asdc.  The spectrum in the
region $t\to+\infty$ is just the free spectrum of the unstable
D-brane, and \sqzed\ is a pure state of open string
excitations. Despite this fact we shall see in section 3 that
\sqzed\ closely resembles a thermal state at temperature $T_H$.
Indeed we shall see that the results of measurements done after
brane creation differ from thermal results by asymptotically
vanishing amounts.

\subsec{Full s-brane modes}

For the full s-brane potential \tis, the Klein-Gordon equation is
\eqn\fulleq{\eqalign{ &(\p^2_t + 2{\l} \cosh t + \omega^2) u=0 .}}
This is a form of Mathieu's equation. We will now summarize a few
useful properties of the solutions -- see e.g. \refs{\gradsh\mclach{--}
\GubserIU} for more
detail. The solutions are generalized Mathieu functions, which can
be written as
\foot{This form for $u$
differs from the standard convention for Mathieu functions.
Our $\tilde \omega$ is related to the standard
characteristic exponent (often denoted $\nu$ in the literature) by
$\nu=-2i\tilde{\omega}$. }
 \eqn\genu{\eqalign{ u=&A e^{-i\tilde{\omega} t}
P(t)+ B e^{i\tilde{\omega} t} P(-t) }} where the function
\eqn\asdx{ P(t) = \sum_{r=-\infty}^{\infty} c_{2r} e^{rt} } is
periodic in imaginary time $P(t)=P(t+2\pi i)$. The constants
$\tilde{\omega}$ and $c_{2r}$ obey complicated recursive formulae
and are typically computed numerically.
Although it is not obvious from \genu,
all solutions $u$ vanish exponentially in the far past and the far
future. This is because open string modes get very massive far
from the interior of the brane.

Solutions to Mathieu's equation are typically classified by their
behavior with respect to imaginary time $\tau=it$. A solution is
bounded on the $\tau$ axis only if $\tilde{\omega}$ is purely
imaginary. In this case the solution is called {\it stable} -
otherwise it is {\it unstable}. Solutions are stable only for
certain regions of the $(\omega^2,{\l})$ plane. For large positive
$\omega^2$ -- the case of interest -- $\tilde{\omega}$ is real and
all solutions are unbounded on the real $\tau$ axis. However,
there is a unique solution that vanishes as $\tau\to+\infty$ --
this is the solution with $B=0$. We will see later that this
solution is naturally associated to the state with no particles in
the interior of the s-brane at $t=0$.

It is useful to assemble Mathieu functions in the form \GubserIU\
analogous to Bessel functions \eqn\asdy{\eqalign{
&J(-2i\tilde{\omega}, t/2) =\equiv e^{-i\tilde{\omega} t} P(t) =
\sum_{n=-\infty}^\infty \phi\left(n-i\tilde{\omega}\right)
e^{(n-i\tilde{\omega})t}, \cr &H^{(1)}(-2i\tilde{\omega},t/2)
\equiv
{J(2i\tilde{\omega},t/2)-e^{-2\pi\tilde{\omega}}J(-2i\tilde{\omega},t/2)
\over \sinh 2\pi\tilde{\omega} }, \cr
&H^{(2)}(-2i\tilde{\omega},t/2) \equiv
{J(2i\tilde{\omega},t/2)-e^{2\pi\tilde{\omega}}J(-2i\tilde{\omega},t/2)
\over -\sinh 2\pi\tilde{\omega} }. }} They have asymptotic
behavior \eqn\asdz{\left.\eqalign{
H^{(1)}(-2i\tilde{\omega},t/2)&\to {\l^{-1/4}\over
\sqrt{\pi}}e^{-\pi\tilde{\omega}}
\exp\left(-{t\over4}+2i\sqrt{{\l}}e^{t/2} - i{\pi\over 4}\right)
\cr H^{(2)}(-2i\tilde{\omega},t/2)&\to {\l^{-1/4}\over \sqrt{\pi}}
e^{\pi\tilde{\omega}} \exp\left(-{t\over4}-2i\sqrt{{\l}}e^{t/2} +i
{\pi\over 4}\right) }\right\} ~~{\rm as }~~ t\to+\infty.}
Mathieu's equation is invariant under $t\to -t$, so
$J(-2i\tilde{\omega},-t/2)$ is also a solution. Under $t\to t+2\pi
i $, $J(-2i\tilde{\omega},-t/2)$ picks up the phase
$e^{-2\pi\tilde{\omega}}$, so it must be proportional to
$J(2i\tilde{\omega},t/2)$ \eqn\asdw{J(2i\tilde{\omega},t/2)=\chi
J(-2i\tilde{\omega},-t/2)} where the proportionality factor $\chi$ is
related to $\phi$ by \eqn\chinu{\chi = {\phi(n+i\tilde{\omega})
\over \phi(-n-i\tilde{\omega})}={\phi(i\tilde{\omega}) \over
\phi(-i\tilde{\omega})}.} The coefficients $\phi(\tau)$ can be
computed using the formula \eqn\phiz{\eqalign{ \phi(\tau)&={1\over
\Gamma(1+\tau+i\omega)\Gamma(1+\tau-i\omega)} \sum_{n=0}^\infty
(-1)^n\l^{2n+\tau} A_\tau^{(n)}, \cr A_\tau^{(0)}&=1, \cr
A_\tau^{({\l})}&=\sum_{p_1=0}^\infty \sum_{p_2=0}^\infty\cdots
\sum_{p_{\l}=0}^\infty a_{\tau+p_1}a_{\tau+p_1+p_2} \cdots
a_{\tau+p_1+\cdots+p_{\l}}, \cr a_\tau &= {1\over (1+\tau+i\omega)
(1+\tau-i\omega) (2+\tau+i\omega) (2+\tau-i\omega)}. }} One can
analyze the behavior of $H^{(i)}(-2i\tilde{\omega},t/2)$ as
$t\to-\infty$ using the relation
\eqn\rel{\eqalign{H^{(1)}(-2i\tilde{\omega},t/2)=&{1\over 2\sinh
2\pi\tilde{\omega}} \cr &\left[ \left(\chi-{1\over\chi}\right)
H^{(1)}(-2i\tilde{\omega},-t/2) + \left(\chi-{e^{-4\pi
\tilde{\omega}}\over\chi}\right) H^{(2)}(-2i\tilde{\omega},-t/2)
\right].}} When $\l$ is small, we can use the expansion
$$
\omega^2=\tilde{\omega}^2+{2\l^2\over 4\tilde{\omega}^2+1} +
{(20\tilde{\omega}^2-7)\l^4\over
2(4\tilde{\omega}^2+1)^3(\tilde{\omega}^2+1)} +\cdots
$$
to compute $\tilde{\omega}$.

\subsec{$|in\rangle_s$ and $|out\rangle_s$ s-brane vacua}

For our full s-brane, the incoming and outgoing positive frequency
wave functions are normalized as \eqn\asd{\eqalign{u^{in}(t) &=
\sqrt{\pi\over 2} (ie^{2\pi\tilde{\omega}})^{1/2}
H^{(1)}(-2i\tilde{\omega},-t/2),~~~~~ \cr u^{out}(t) &= \sqrt{\pi\over
2}
(ie^{2\pi\tilde{\omega}})^{-1/2}H^{(2)}(-2i\tilde{\omega},t/2).}}
The in (out) vacuum, which has no incoming (outgoing) particles,
is defined by the condition
\eqn\zjs{a^{in}|in\rangle_s=0=a^{out}|out\rangle_s.} The relation
between in and out modes is determined from \rel\ to be
\eqn\asd{\eqalign{ u^{in}(t)&={1\over 2\sinh2\pi\tilde{\omega}}
\left[ i\left(e^{2\pi \tilde{\omega}} \chi-{e^{-2\pi
\tilde{\omega}}\over \chi} \right)u^{out}(t) + \left(\chi -
{1\over\chi}\right) u^{out*}(t) \right]\cr &=\alpha
u^{out}(t)+\beta u^{out*}(t), }} where $\alpha,\beta$ are the
Bogolubov coefficients \eqn\asd{\alpha = {i\over
2\sinh2\pi\tilde{\omega}}\left(e^{2\pi \tilde{\omega}}
\chi-{e^{-2\pi \tilde{\omega}}\over \chi} \right),~~~~\beta =
{1\over 2\sinh2\pi\tilde{\omega}}\left(\chi - {1\over\chi}\right).
} Although the dependence of $\tilde{\omega},\chi$ on
$\omega,{\l}$ is in general quite complicated, it is known that
either $e^{2 \pi \tilde{\omega}}$ is real and $\chi$ is of unit
modulus, or $\chi$ is real and $e^{2\pi\tilde{\omega}}$ is of unit
modulus. It follows that the Bogolubov coefficients
satisfy the unitarity condition $|\alpha|^2-|\beta|^2=1$. In the
case of interest, $\omega$ is large and ${\l}\ll\omega$, we have
\eqn\omapprox{\tilde{\omega} = \omega \left[1+O({{\l}^2/
\omega^4})\right].} Using \chinu\ and \phiz, we obtain the
expansion for $\chi$ \eqn\asd{\chi =
{\Gamma(1-2i\omega)\over\Gamma(1+2i\omega)}\l^{2i\omega} \left[
1+O({\l}^2/\omega^2) \right].}  To leading order in
$O({\l}^2/\omega^2)$, the Bogolubov coefficients are
\eqn\bogofull{\eqalign{\alpha =& {\sin(\theta+2\pi i\omega)\over
\sinh2\pi\omega},~~~~ \beta = -i{\sin\theta\over\sinh2\pi\omega},
~~~~~
e^{i\theta}=\l^{-2i\omega}
{\Gamma(1+2i\omega)\over\Gamma(1-2i\omega)}. }} Sub-leading terms
can be computed  order by order in $1/\omega^2$ using the methods
of \GubserIU, although we shall not need them here.

\subsec{The $|0\rangle_s$ Euclidean s-brane vacuum  }

If we take $\lambda \to 0$ there is a long region around $t=0$,
of duration $\ln \lambda $, in which the interaction can be
neglected and we just have an ordinary unstable brane. There is
then a natural $|0\rangle_s$ vacuum in which there are no
particles present at $t=0$.  This will later be identified with a
vacuum obtained by analytic continuation from Euclidean space. It
is associated with the wave functions \eqn\uzero{u^0(t) =
\sqrt{4\pi \chi \over \sinh2\pi\tilde{\omega}}
J(-2i\tilde{\omega},t/2).} Using the relation $J=\half
[H^{(1)} + H^{(2)} ]$, $u^0(t)$ can be expressed in terms of
$u^{out}(t)$ as \eqn\asd{\eqalign{u^0(t)&=\sqrt{\chi\over{2\sinh
2\pi\tilde{\omega}}} \left[ e^{\pi\tilde{\omega} + i{\pi\over
4}}u^{out}(t) + e^{-\pi\tilde{\omega} -i {\pi\over 4}}u^{out*}(t)
\right] \cr &= a^*u^{out}(t)-bu^{out*}(t) }} and in terms of
$u^{in}(t)$ as \eqn\asd{\eqalign{ u^0(t)&={1\over\sqrt{2\chi\sinh
2\pi\tilde{\omega}}} \left[ e^{\pi\tilde{\omega} - i{\pi\over
4}}u^{in}(t) + e^{-\pi\tilde{\omega} + i{\pi\over 4}} u^{in*}(t)
\right] \cr &= au^{in}(t)-b^*u^{in*}(t). }} The Bogolubov
coefficients relating the $|0\rangle_s$ vacuum to the $|in\rangle_s$
and $|out\rangle_s$ vacua are given by \eqn\bogthree{\eqalign{ 
a&= {e^{\pi\tilde{\omega} -i {\pi\over 4}} \over
\sqrt{2\chi\sinh2\pi\tilde{\omega}}},~~~~~b = -
e^{-\pi\tilde{\omega} -i{\pi\over 4 }}\sqrt{\chi\over
{2\sinh2\pi\tilde{\omega}}}. }}

In the limit $\omega,\omega/\l\gg 1$, the s-brane wave functions may
be understood in terms of a flux matching procedure.
\foot{A similar procedure was used to study scattering of scalar
fields by a D3 brane in \KlebanovKC.}
In this limit the Klein-Gordon equation \fulleq\ describes brane creation
\halfeq\ in the far past and brane decay in the far future, separated
by a long region near $t=0$ where the interaction is negligible.
Approximate solutions to \fulleq\ may be found by matching
wave functions of brane creation with wave functions of brane decay
across the region $t \sim 0$.
In particular, $u^{in}$ and $u^{out}$
look like the half s-brane solutions $u^{in}_-$ and $u^{out}_+$ in
the past and future, respectively.  The wave function
$u^0$ looks like an ordinary
plane wave solution in the interior of the s-brane, $u^{out}_-$ in the
far past and $u^{in}_+$ in far future.
This may be verified by noting that to lowest order in $1/\omega^2$ the
Bogolubov coefficients \bogthree\ are given by \candd.

For transitions to and from the $|0\rangle_s $ vacuum, we find
that the particle creation rate is governed by the ``thermal"
factors : \eqn\gammais{\eqalign{ \gamma_{0\to in}& = b/ a^* =i
e^{-2\pi\tilde{\omega}} ,~~~~~~~~~~~~~~~ \gamma_{in\to 0} =
-b/a = e^{-2\pi\tilde{\omega}-i\tilde{\theta}} \cr
\gamma_{0\to out}& = b^*/a=-i e^{-2\pi\tilde{\omega}}
,~~~~~~~~~~~~~~~ \gamma_{out\to 0}=-b^*/a^*=
e^{-2\pi\tilde{\omega}+i\tilde{\theta}} ,}} where
$e^{-i\tilde{\theta}}=\chi$. In the limit of large $\omega$ and
$\omega/\lambda\gg 1$,
$\tilde{\omega},\tilde{\theta}\to\omega,\theta$ and we find the
same particle creation rates as for the half s-branes. We see that the
particle occupation numbers in the far past and future are both
thermal at temperature $T=1/4\pi$.
The $|0\rangle_s$ vacuum is special in this respect.

The $|0\rangle_s$ state is also special because it is time
reversal invariant and is the natural state defined by analytic
continuation 
 from Euclidean space. This latter property can be readily seen 
from the fact that  it is the unique
state whose modes are bounded on the positive Euclidean axis, as
described in section 2.2. It is also the only state whose positive
and negative frequency modes do not mix under imaginary time
translation $t\to t+2\pi i$.
A general solution of the wave equation has the form \eqn\asd{ u=A
J(-2i\tilde{\omega},t/2) + B J(-2i\tilde{\omega}, -t/2) .} 
For the $|0\rangle$ vacuum $B$ vanishes, and the 
solution transforms as \eqn\transl{u^0(t+2\pi i)= e^{2\pi
\tilde{\omega}} \ u^0(t)} under imaginary time translation.
In all other states both $A$ and $B$ are non-zero, so this
procedure mixes $u$ with $u^*$.  When $\omega$ is large and $\omega/\l\gg
1$, $\tilde{\omega}$ approaches $\omega$ and \transl\ is the usual
imaginary time translation condition exhibited by plane waves in flat
space.  Moreover, in our choice of solutions
\uzero\ we have chosen
the phase of $A$ so that $u^0$ has the usual time
reversal behavior \eqn\asd{u^0(-t)=u^0{}^*(t).}

\subsec{More vacua}

The wave equation \fulleq\ is second order,
so for any given momentum
$\pb$ there is a one complex parameter family of normalized,
positive frequency solutions $u(t)$.  We can write these solutions  as
linear combinations \eqn\asd{ u^\a = {1\over
\sqrt{1-e^{\a+\a^*}}}\left(u^0+e^{\a} u^0{}^*\right) } where $\a$
is a function of $\pb$ with negative real part.  This choice of
modes defines a state $|\a\rangle$ for every such function
$\a(\pb)$. The state is invariant under spatial translations when
$\a$ is a function of $p$ only.
This is a very large family of vacua, which includes the
$|in\rangle_s$, $|0\rangle_s$ and $|out\rangle_s$ states.
Of course, one can define families of vacua for brane creation
and brane decay as well.

In order to constrain $\a$ further, one needs to demand some other
symmetry.  One such symmetry is time reversal invariance -- this
restricts us to states such as $|0\rangle_s$ and ${1\over 2}
(|in\rangle_s + |out\rangle_s)$. It is also natural to demand that the
short distance structure of the vacuum be the same as the $|0\rangle_s$
vacuum.  This restricts $\a(p)$ to vanish
sufficiently quickly at large $p$.

\newsec{S-brane thermodynamics }

In this section we study the thermal properties of  s-branes.
In the first subsection we will consider the response of a
monopole Unruh detector coupled to $\phi$.  We will show that in
the $|in\rangle_-$ and $|out\rangle_+$ vacua the detector response
is thermal, but that in other states the detector response is
non-thermal. Next, we will demonstrate that in the $|in\rangle_-$
vacuum the correlators of the theory become exactly thermal up to
corrections that vanish in the far future.  Likewise, the
$|out\rangle_+$ vacuum looks thermal in the far past. Similar
results pertain to the full s-brane, but we will not work out the
details here.

\subsec{S-branes at finite temperature}

The s-branes described by \tpp-\tis\  are highly
time-dependent configurations. Temperature is an equilibrium
(or at best adiabatic) concept, so it usually does not make sense to
put a time dependent configuration at finite temperature unless
the inverse temperature is much lower than the scale of time variation.
Thus it would seem to be impossible to study s-branes at 
string-scale temperatures.
However, certain special properties of s-branes make this possible,
as we will now demonstrate for a scalar field obeying \zasd. In section 5 we will see that the finite
temperature states in string theory have a natural boundary state
description. 

Let us first consider the
 construction of Green functions by analytic continuation from
 Euclidean space. In Euclidean space, parameterized by $\tau=it$, the formulae
 \tpp-\tis\ for $m^2(\tau)$ have the special property that they
 are periodic under $\tau \to \tau +2\pi$. It is therefore
 possible to identify Euclidean space so that $\tau=\tau +2\pi n$ for any
 integer $n$, and compute the Green function on the identified
 space. After continuing back to Minkowski space, the resulting
 Green function $G_{2\pi n}(\vec
 x, t;\vec x',t')$ will be periodic under $t \to t+2\pi i n$
and $t' \to t'+2\pi i n$. This is
 a thermal Green function at temperature $T={1 \over 2\pi n}$.

The thermal Green function so obtained can also be understood as a
two point function in a certain mixed state.  We will consider the
case of brane decay, although a similar discussion will apply to brane
creation and to the full s-brane.  For $t,t' \to
-\infty$, $G_{2\pi n}(\vec x, t; \vec x',t')$ is clearly the usual
thermal Feynman Green function which is given by (suppressing the
time-ordering)\eqn\rtf{\eqalign{G_{2\pi n}(\vec x, t;\vec x', t')&=\sum_j
e^{-2\pi n E_j}\langle E_j|\phi(\vec x',t')\phi(\vec x,t)|E_j\rangle
\cr &={\rm Tr}\rho_{n} \phi(\vec x',t')\phi(\vec x,t).}}
Here we have defined the density matrix
\eqn\fza{\rho_n =C_ne^{-2\pi n H(-\infty)}} where $C_n$ is a normalization
constant and the time dependent
Hamiltonian is \eqn\jji{H(t)=\int_\Sigma d^dx\bigl((\dot \phi)^2
+(\nabla \phi)^2+e^{2t}\phi^2\bigr)+N.} In this expression,
$N$ is a (time-independent) normal ordering constant and $\phi$ is to be
expanded in terms of time-independent creation and annihilation operators
as
\eqn\expp{\phi=\sum_{\vec p} \bigl(\psi^{in}_{\vec p} a^{in}_{\vec p} +\psi^{in*}_{\vec p
}a_{\vec p}^{in\dagger}\bigr)=\sum_{\vec p} \bigl(\psi^{out}_{\vec p} a^{out}_{\vec p} +\psi^{out*}_{\vec p
}a_{\vec p}^{out\dagger}\bigr).}
The Hamiltonian $H(t)$ obeys
\eqn\tfs{i[H(t), \phi(t)]=\p_t\phi(t),~~~~~~~i[H(t),
\p_t\phi(t)]=\p_t^2\phi(t)}
so that the path-ordered operator
\eqn\fcv{U(t_2,t_1)=P\bigl[e^{i\int_{t_1}^{t_2} H(t)dt} \bigr]}
generates time evolution on $\phi$ \eqn\tevl{\phi(\vec x,
t_2)=U(t_2,t_1)\phi (\vec x, t_1)U(t_1,t_2).} However, because of the explicit time dependence in \jji,
$U$ does not generate evolution of $H(t)$.

In fact \rtf\ turns out to be the desired Green function even for
finite $t,t'$, and \fza\ can be viewed as the exact
Heisenberg-picture density matrix.  First, it can be shown from 
properties of Bessel functions that
\eqn\ftxz{\phi(\vec x, t+2\pi i n
)=e^{-2\pi n H(-\infty)}\phi(\vec x,t)e^{2\pi n H(-\infty)},}
\eqn\txz{\p_t\phi(\vec x, t+2\pi i n )=e^{-2\pi n
H(-\infty)}\p_t\phi(\vec x,t)e^{2\pi n H(-\infty)}.} Note that
shifts in $t\to t+2\pi i$ are generated by $H(-\infty)$ for {\it any} $t$.
This implies \eqn\xxb{U(t,t+2\pi i n)=U(-\infty, -\infty +2\pi i
n)={\rho_n \over C_n} .} It then follows that \rtf\ obeys
\eqn\dsc{G_{2\pi n}(\vec x, t+2\pi i n ; \vec x',t')=G_{2\pi
n}(\vec x, t; \vec x',t'-2\pi i n)=G_{2\pi n}(\vec x', t'; \vec
x,t),} as expected for a thermal Green function.

It is possible to define a time-dependent
Schroedinger picture density matrix
\eqn\fxcb{\rho_n(t)= C_n U(-\infty, t+2\pi i n)U(t, -\infty).}
One can then compute correlators at equal (finite) times
by inserting the Schroedinger-picture operators
$\phi(\vec x,-\infty)$:
\eqn\dsxb{G_{2\pi
n}(\vec x, t; \vec x',t)={\rm Tr}\rho_n(t)\phi(\vec x,-\infty)\phi(\vec
x',-\infty).}
The Heisenberg density matrix is related to the Schrodinger density
matrix by
\eqn\asd{
\rho_n=\rho_n(-\infty).
}

We note that the thermal density matrix $\rho_n$ can not be
exactly identified with any of the vacua of the preceding section,
all of which are pure states by construction. Despite this fact we
will see below that in some cases it becomes difficult to
distinguish pure and mixed states.

To summarize, we have seen that, despite the time dependence of
the background,  it is mathematically possible to define mixed
states which approaches the standard thermal vacuum at past
infinity while retaining the thermal periodicity at all times.

\subsec{Unruh detection }

In the previous subsection, a sequence of mixed thermal states
were described with temperatures $T={1 \over 2\pi n}$. On the
other hand, in the previous section we saw that various s-brane
vacua -- despite being pure states -- have some mysterious "thermal"
behavior. In this subsection we will clarify this by showing that
a particle detector in these vacua will respond as if it is in a
mixed thermal state with temperature $T={1 \over 4 \pi}$. In the
next subsection we will show that the pure state correlators can
asymptotically approach those of the thermal state.

Let us imagine coupling some detector to the field $\phi$ via a
monopole interaction term $\int dt \ O(t) \phi(t)$.  Here $O(t)$
is a hermitian operator that acts on the Hilbert space of the
detector, which we will assume is spanned by some discrete,
non-degenerate set of energy eigenstates $|E_i\rangle$. We also
assume that the detector is stationary and that the detector
Hamiltonian is time independent.  We can now calculate the
probability that the detector will jump from a state with energy
$E_i$ to one with
energy $E_j$. 
To first order in perturbation theory it is
(a recent discussion appears in section 3.2 of \SpradlinPW)
\eqn\pij{ P_{i\to j} =
|\langle E_i|O(0)|E_j\rangle|^2 \int dt \int dt'
e^{-i\Delta E (t-t')} G(t,t') } where $\Delta E=E_i-E_j$ and
$G(t,t')$ is the Wightman two point function of the scalar field in a
particular vacuum state. The detector response is thermal at
temperature $T$ if the probability amplitudes \pij\ obey the
detailed balance condition \eqn\prat{ {P_{i\to j} \over P_{j\to i}
}= e^{-\Delta E / T} .} For time translation invariant theories
the green function depends only on $t-t'$ and the double integral
\pij\ is infinite.  One can then factor out the $\int d(t+t')$ to
get a finite expression for the transition rate per unit time
${\dot P}_{i\to j}$. For s-branes the Green function $G(t,t')$ is
not time translation invariant and the calculation is more
complicated.

In fact, for the full s-brane expression \pij\ -- the {\it total}
probability integrated over all time -- is finite. This is because
the Green function $G(x,y)$ solves the wave equation in both
arguments, so that as $t\to\pm\infty$, $G(t,t')\sim e^{\mp t/4}$
vanishes rapidly.  This holds for both $t$ and $t'$, so the double
integral $\pij$ converges. This is true for any vacuum state of
the full s-brane. This behavior has a natural physical
interpretation: in the far past and far future the open string
states become very massive and cannot couple to the detector.

For brane creation, the Green function $G(t,t')$ does not fall off
exponentially as $t\to+\infty$, so that the integrated probability
amplitudes \pij\ are infinite.  At this point one approach is to
consider the transition rates ${\dot P}_{i\to j} (t)$, which are
finite and time dependent. We will take a different point of view,
however, and show directly that the ratio \prat\  converges to a finite
(and interesting) answer.

In order to evaluate \pij, we rewrite the Green function in a
particular vacuum as \eqn\asd{\eqalign{
G(x^0,\xb;y^0,\yb)&=\langle| \phi(x^0,\xb)\phi(y^0,\yb)|\rangle
\cr &=\int d^{d-1}p \ e^{-i\pb\cdot(\xb-\yb)} u^* (x^0) u (y^0) }}
where $u$ are the ($p$-dependent) positive frequency modes. Then
\pij\ becomes \eqn\pijj{\eqalign{ P_{i\to j}=& \int d^{d-1}\pb
\ |{\tilde u} (E_i-E_j)|^2 }} where ${\tilde u}$ are the Fourier
transformed modes. For the $|out\rangle_+$ vacuum of brane decay we can
evaluate the Fourier integral \foot{ As usual, the integral
converges only after we deform the contour to give $E$ a small
positive imaginary part, $E\to E+i\epsilon$. } \eqn\asd{\eqalign{
{\tilde u}^{out}_+ (E)&=\int e^{-iEt} u^{out}_+(t) \ dt\cr
&=\sqrt{i\over 8 \pi} e^{-\pi E } \l^{iE}
{\Gamma(-{i}(\omega+E))\Gamma({i}(\omega-E))} .}} This has the
form \eqn\asdx{ {\tilde u}^{out}_+ (E) = i^{1/2} e^{-\pi E} x(E) }
where $x(-E) = x^*(E)$.  The norm is \eqn\asdy{\eqalign{ |{\tilde
u}^{out}_+ (E)|^2&= e^{-2\pi E} \left[x x^*\right]
,}} where the term in the square brackets is invariant under
$E\to-E$. From this fact and expression \pijj\ it follows that the
transition probabilities in the $|out\rangle_+$ vacuum
satisfy the detailed balance condition
\prat\ with characteristic temperature $T_H={1\over 4\pi}$. \foot{ As
mentioned above, these transition probabilities are infinite, so
in evaluating \pij\ it is necessary to regulate the divergent
momentum integrals by, e.g. imposing some cutoff. In evaluating
the ratio \prat\ one can safely take this regulator to infinity.
}
The $|in\rangle_-$ vacuum for brane creation  is related to the
$|out\rangle_+$ vacuum
for brane decay  by time reversal, so it has identical transition
probabilities $P_{i \to j}$ and thermal detector response.

In fact, the $|in\rangle_-$ and $|out\rangle_+$ vacua are very special
in this respect.  For example, we could
calculate the transition probabilities in the $|in\rangle_+$
vacuum by inverting the Bogolubov relation \bogone.
This gives \eqn\asdz{\eqalign{
{\tilde u}^{0}_p (E)
&= c^* {\tilde u}^{out}_p (E) - d {\tilde u}^{out}_p{}^* (-E)
\cr& = - 2 i^{3/2} d e^{\pi\omega} \cosh{\pi}(\omega-E) x
.}} 
The norm is \eqn\asda{ |{\tilde u}^{0}_p (E)|^2 =
\cosh^2{\pi}(\omega-E) \left[4 |d|^2 e^{2\pi\omega} x
x^*\right] } where again the quantity in the square brackets is
invariant under $E\to-E$. In this case it is clear that the
transition probabilities are not thermal. It seems likely that for
{\it any} vacuum state that is not $|out\rangle_+$ or
$|in\rangle_-$, the Bogolubov transformation will give terms
proportional to $e^{\pi E}$ so that the detector response is
non-thermal.

\subsec{Thermal correlators }

We will now demonstrate that the correlators of the $|in\rangle_-$
vacuum are asymptotically thermal in the far future. The theory is
free, so it suffices to consider the two point function $G(x,y)$.
For brane creation, we have \eqn\uinis{ u_-^{in} = a ( u_-^{out} -
e^{-2\pi\omega +i\theta} u_-^{out}{}^* )}
so the $|in\rangle_-$ Green function is \eqn\ginis{\eqalign{
G_-^{in}(x,y) =& \int d^{d-1}\pb \ u^{in}(x) u^{in}{}^*(y) \cr =&
\int {d^{d-1}\pb\over 1-e^{-4\pi\omega}} e^{i\pb\cdot(\xb-\yb)}
\left[ ( u_0(t) u^*_0(t') + e^{-4\pi\omega} u^*_0(t) u_0(t'))
\right.\cr &~~~~~~~~~~~~~~~~~~~~~~~~~~~~~~~~~~~~~~~ \left. + e^{-2\pi\omega} (e^{i\theta} u_0(t) u_0(t')
+ c.c.) \right] .}}
In the far future, $u_-^{out}$ approaches a positive frequency
plane wave plus corrections exponentially small in $t+t'$.
In this limit the
term on the second line of \ginis\ becomes a function of $t-t'$ only,
and approaches the usual (constant mass) thermal Green function 
at temperature $T_H={1 \over 4\pi}$
\eqn\asd{\eqalign{ G^T(x,y) = \int
{d^{d-1}\pb\over 2\omega} \left( {e^{i\pb\cdot(\xb-\yb) - i\omega
(t-t')}\over 1-e^{-4\pi\omega}} -  {e^{i\pb\cdot(\xb-\yb) +
i\omega (t-t')}\over 1-e^{4\pi\omega}}\right) .}}
In the far future the third line
of \ginis\ depends on $t+t'$ rather than $t-t'$, and
gives a contribution to the Green function
\eqn\gextra{ \int {d^{d-1}\pb\over 2\omega} {2\over
\sinh2\pi\omega} \left( e^{i\theta} e^{i\pb\cdot (\xb-\yb)
-i\omega(t+t')} + c.c. \right) } plus exponentially small
corrections.  In the limit $t,t'\to+\infty$ this contribution
vanishes as $(t+t')^{-(d-1)/2}$.
In fact, when $\omega^2= p^2$ (i.e. the field becomes massless in the
far future) the integral \gextra\ can be converted into a contour integral
and shown to vanish exponentially in $t+t'$.

We conclude that in the far future the pure state $|in\rangle_-$
correlators become thermal plus asymptotically vanishing
corrections. Likewise the $|out\rangle_+$ correlators become
thermal in the far past.

\newsec{Long-distance s-brane effective field theory }

The dynamics of ordinary D-branes are described at low energies by
a long-distance effective field theory. This field theory is of
much interest in the  understanding of holographic bulk-brane
duality. One would like to know if there is a similar
long-distance field theory for s-branes, which would be a
candidate holographic dual for an appropriate bulk string
cosmology.

The first task is to define the notion of an effective field
theory for branes with a spacelike orientation. This is best
understood in terms of correlators. We define the long-distance
effective field theory
on the p+1-dimensional sp-brane as the p+1-dimensional
Euclidean field theory that reproduces
the long-distance correlators.
To be specific, let us
consider an s2-brane which is real codimension one in four
spacetime dimensions. A massless field confined to the s2-brane should
have a correlator that falls off like ${1 \over r}$. This is
quite distinct from massless correlators at
spacelike-separated points in the ambient
four-dimensional spacetime, which fall off like $1 \over r^2$.

Consider a scalar on the full s2-brane with four-dimensional wave
equation: \eqn\weqa{\partial^\mu \partial_\mu \phi -(2{\l} \cosh t
+\omega^2)\phi=0,} as in \fulleq. In the far future and far past,
$\phi$ is very massive, and correlators fall off exponentially
with the spatial separation. Hence if there are any massless
excitations they will be confined to the s2-brane world-volume near
$t=0$ where $\phi$ is light.

In trying to compute the equal-time correlators near $t=0$, we
immediately encounter a puzzle. Such correlators depend on the
choice of quantum state for the field $\phi$. Indeed, given any
set of equal time correlators $\Delta(\vec x,\vec y)$ there exists
a quantum state $\Psi[\phi ]=exp[-{ 1 \over 4} \int \int  \phi(\vec x)
\Delta^{-1}(\vec x,\vec y) \phi(\vec y)]$ that reproduces them.

In order to determine the long-distance effective field
theory we must therefore first specify a quantum state.  One possibility is to
take the $|0\rangle_s$ state which has no particle flux at $t=0$.
Spacelike correlators at $t=0$ in this state fall off exactly as
they would in Minkowski space, $i.e.$ as ${1 \over r^2}$. So this
does not lead to a low energy effective field theory confined to
the s-brane world-volume.  As we have seen,
a natural state for an s-brane is the thermal state
at temperature $T={1 \over 2\pi n }$. Such thermal states
plausibly approximate the quantum states of open strings on
an s-brane created from incoming closed string excitations.
The $t=0$
correlators in these states indeed fall off as ${1 \over r}$, indicating
massless modes are confined to the s-brane.  This can be most
easily seen from the Euclidean construction of the correlators on
$R^3\times S^1$. At distances large compared to the radius of the
$S^1$ (i.e. large compared to the inverse temperature) there is an effective
compactification from four to three (Euclidean) dimensions, and so
the effective correlators are three-dimensional. Massless modes of
the three dimensional effective theory arise as usual from
compactification zero modes. From the Minkowskian perspective, the
mixed thermal state has excited components which carry more
spatial correlations than the vacuum.

The Euclidean zero mode equation following from \weqa\ is
\eqn\wqa{ \partial_\tau^2 \phi -(2{\l} \cos \tau
+\omega^2)\phi=0,} where $\tau \sim \tau+2\pi n$. Whether or not
there is a zero mode for a given temperature depends on the
precise value of the mass parameter $\omega^2$ \wqa, which has to be
fine-tuned to get an exact zero mode. This special value might
arise as a consequence of Goldstone's theorem or other
symmetry considerations.

In conclusion, the s2-brane has a naturally associated
three-dimensional Euclidean effective field theory given roughly
by the high-temperature limit of the theory on the
four-dimensional unstable brane. The full determination of such an
effective theory for stringy s-branes is beyond the scope of the
present work, although a few preliminary comments are made in our
concluding discussion.

\newsec{Thermal boundary states}
 We have seen that natural quantum states for s-branes
are mixed thermal states at temperature $T={1 \over 2\pi n}$.
In this section we will construct the exact CFT boundary state
whose worldsheet correlators give the thermal spacetime Green functions,
generalizing the zero-temperature construction of \SenNU.
Ghost and spacelike components of the boundary state are suppressed, but are
similar to those in \SenNU.
\subsec{Zero modes and winding sectors}
 The boundary
state $| B\rangle$ for the conformally invariant boundary
Sine-Gordon theory
\eqn\cklma{S={1\over 2 \pi} \int_{\Sigma}
\partial \phi \bar
\partial \phi
  +{ \lambda \over 2 } \int_{\partial \Sigma}  \big(  e^{i \phi}+ e^{-i\phi}\big)}
was found using the bulk $SU(2)$ current algebra in \CallanUB\
(see also \refs{\PolchinskiMY , \GreenGA , \GaberdielZQ}). Here $\phi$ is
Euclidean time, which will later be
analytically continued to the Lorentzian world sheet field $X^0=i\phi$.
The boundary state corresponding to temperature $T= {1 \over 2\pi
n}$ arises when one identifies \eqn\idth{\phi \sim \phi +2\pi n.}
This identification restricts the left and right moving momenta to be
\eqn\momare{
(p_L,p_R)=({p \over n}+{wn}, {p \over n}-wn)
}
where $p$ and $w$ are integers.
In this case the thermal boundary state is simply \eqn\cklmb{ |
B_n\rangle = P_n \ e^{2 \pi i \lambda J_1 }|N\rangle.}
Here
\eqn\nbst{ |N\rangle=2^{-1/4}\sum_{j,m}| j;m,-m\rangle\rangle}
is the standard $SU(2)$ Neumann boundary state and
$| j;m,-m\rangle\rangle$ is the Ishibashi
state associated with the SU(2) primary field $|j;m,-m\rangle$.
The SU(2) rotation $J_1$ acts only on
right-movers, and $P_n$ is the
projection operator onto the allowed sub-lattice defined by \momare.
For $n=1$, $P_1$
is the identity. In the non-compact case, $n \to \infty$ and
$P_\infty$ projects onto $p_L=p_R$.

Let us now consider the part of the boundary state which involves
no oscillators. There are four such terms for every $j$,
namely $| j;\pm j,\pm j \rangle$.
Now, since
\eqn\asd{\eqalign{
\langle j;j,j|P_n=\langle j;j,j|P_\infty,~~~~~~
\langle j;-j,-j|P_n=\langle j;-j,-j|P_\infty}}
for any $n$,
the $p_L=p_R$ components of the state are the same as in \SenNU.
For real $\lambda$ they may be written \eqn\dfz{\bigl[1+2\sum_{j
\neq 0}(-\sin (\pi \lambda ))^{2j}\cos (2j \phi(0)) \bigr]|
0\rangle = {\cos ^2 (\pi \lambda) \over 1+\sin^2( \pi
\lambda)+2\sin(\pi \lambda) \cos \phi(0) }| 0\rangle.} For
finite $n$, there are also terms with $p_L=-p_R =2nj \neq 0$.
These are related to the $p_L=p_R$ terms by the rotation $e^{{i \pi
}J_1}$, which corresponds to a shift of $\lambda$ by $\half$.
Hence in addition to \dfz, $| B_n \rangle$ has a winding-sector
component \eqn\dfxz{2\sum_{j \neq 0}(\cos \pi\lambda)^{2nj}\cos
(2nj\tilde\phi(0))| 0\rangle = {2\cos^n(\pi \lambda) \cos(n
\tilde \phi(0))-2\cos^{2n} (\pi \lambda) \over 1+\cos^{2n}( \pi
\lambda)-2\cos^n(\pi \lambda) \cos(n \tilde \phi(0))   }|
0\rangle.} In this expression
$\tilde \phi(z,\bar z)=\half(\phi(z)-\phi(\bar z))$
is the T-dual of $\phi$.

The continuation $\phi \to -iX^0$
of these expressions to the timelike theory
\eqn\cklmx{S=-{1\over 2 \pi} \int_{\Sigma}
\partial X^0 \bar
\partial X^0
  +{ \lambda } \int_{\partial \Sigma} \cosh X^0 }
is straightforward. The theory \cklmx\ contains
the currents\foot{ In our conventions
$X(z,\bar z)=\half
(X(z)+X(\bar z))$,  $X(z)X(w)\sim 2\ln (z-w)$ and $\alpha'=1$.}
\eqn\rftg{j_\pm(z) =e^{\pm
X^0(z)},~~~~~~j_3(z)={1 \over 2}\p X^0(z) ,} which
generate the usual level
one SU(2) current algebra.
 Note however that with the
standard norm $X^{0\dagger}=X^0$, $j_3$ is anti-hermitian while
$j^\pm$ are both hermitian. Nevertheless the charges
\eqn\cdgf{J_\pm=\oint {dz \over 2\pi i} j_\pm(z), ~~~~~~J_3=\oint
{dz \over 2\pi i} j_3(z), } obey the usual commutation relations
\eqn\fcv{[J_-,J_+]=-2J_3,~~~~~~[J_3,J_\pm]=\pm J_\pm .}
States can therefore be characterized by their SU(2) representations.
Under $\phi \to -iX$, $J_k \to J_k$ and hence $| j;
m,m'\rangle\rangle \to | j; m,m'\rangle\rangle$. Therefore the
Sine Gordon boundary state written in the form \cklmb\ can also be
viewed as a boundary state for the timelike theory \cklmx.
Expressions \dfz\
and \dfxz\ become \eqn\fz{\eqalign{\bigl[1+2\sum_{j \neq 0}(-\sin
(\pi \lambda ))^{2j}& \cosh (2j X^0(0)) \bigr]| 0\rangle \cr &=
{\cos^2 (\pi \lambda) \over 1+\sin^2( \pi \lambda)+2\sin(\pi
\lambda) \cosh X^0(0) }| 0\rangle}} and
\eqn\fxz{\eqalign{\sum_{j \neq 0}(\cos \pi \lambda)^{2nj}&\cosh
(2nj\tilde X^0(0))| 0\rangle  \cr &= {2\cos^n(\pi \lambda)
\cosh (n \tilde X^0(0))-2\cos^{2n} (\pi \lambda)\over 1+\cos^{2n}(
\pi \lambda)-2\cos^n(\pi \lambda) \cosh(n \tilde X^0(0))   }|
0\rangle.}} Here $\tilde X^0(z,\bar z)=
\half(X^0(z)-X^0(\bar z))$.

\subsec{The superstring} In this
subsection we sketch the finite-temperature generalization for the
superstring of the zero-temperature boundary state of \SenIN. For
the superstring, instead of the boundary interaction \sdef, one
has \eqn\supers{\lambda\int d\tau\psi^0 \sinh {X^0 \over
\sqrt{2}}\otimes \sigma_1.} We follow the notation of
\refs{\SenMG,\SenIN, \okd} in which $\sigma_1$ acts on Chan-Paton
factors, and \supers\ arises from a tachyon field proportional to
$\cosh {X^0 \over \sqrt{2}}$. After integrating out $\psi^0$ one
obtains a boundary interaction of the form \tis. This interaction
is invariant under \eqn\dsxcl{X^0\to X^0 + 2\sqrt{ 2}\pi i.} We
can therefore consider thermal boundary states at temperatures
\eqn\faz{T={1 \over 2 \sqrt {2}\pi n}.} This corresponds to the
superstring Hagedorn temperature $T_H={1 \over 2 \sqrt {2}\pi }$
for the minimal value $n=1$, rather than the $n=2$ we encountered
in the bosonic string. The bosonic zero mode part of the boundary
state with no winding is then \eqn\fzs{ {1-\sin^4 (\pi \lambda)
\over 1+\sin^4( \pi \lambda)+2\sin^2(\pi \lambda) \cosh
(\sqrt{2}X^0(0)) }| 0\rangle.} The winding component is
\eqn\fxsz{ {2\cos^{2n}(\pi \lambda) \cosh (\sqrt{2} n \tilde
X^0(0))-2\cos^{4n} (\pi \lambda)\over 1+\cos^{4n}( \pi
\lambda)-2\cos^{2n}(\pi \lambda) \cosh(\sqrt{2} n \tilde X^0(0))
}| 0\rangle.} These components are similar to those of the
bosonic string, with the replacement $\sin \pi \lambda \to \sin^2
\pi \lambda$ and the factors of $\sqrt 2$. In computing the
fermionic components, twisted boundary conditions around the
thermal circle must be taken into account.

\newsec{The $\lambda=\pm \half$ sD-brane limit}
In this section we will consider the very interesting limits
$\lambda \to \pm \half$. In \okd\ it was shown that in a certain
$\lambda \to -\half$ limit, the general bosonic s-brane boundary
state \cklmb\ (at zero temperature) reduces to the boundary state
of \mgas\ which imposes a Dirichlet boundary condition $X^0=0$ in
the time direction. In other words $\lambda=-\half$ is an
sD-brane. The relation to sD-branes follows
immediately from the fact that the $\lambda=\pm \half$ boundary
interaction is an $SU(2)$ rotation by $\pi$ which transforms a
Neumann boundary state into a Dirichlet state.

On the other hand, in \SenNU\ it was shown that in a certain
$\lambda \to \half$ limit, the general s-brane boundary state in
some sense reduces to nothing -- i.e. in this limit
there is no brane present at
all.\foot{As can be seen from \fzs, the superstring case for both
$\lambda\to \half$ and  $\lambda\to- \half$ is qualitatively
similar to the $\lambda\to \half$ bosonic case.} In fact, we will
see that the limiting closed string configuration
is not unambiguously
determined from the boundary state for any value of $\lambda$.
Additional boundary conditions
on the fields are needed. In general the limit is not the trivial
one described in \SenNU, but rather is a very special type of
s-brane configuration described by spacelike Dirichlet branes
located on the imaginary time axis.

In the next subsection, we will describe this ambiguity in the
limit $\lambda\to{1\over 2}$. In section 6.2 we will derive  the
linearized RR-field sourced by an sD-brane and see that it carries
a non-trivial s-charge for all $\lambda$. In 6.3 the long range 
graviton and dilaton fields of an s-brane are computed. In 6.4 we determine
the force between an ordinary D-brane and a $\lambda=\half$
sD-brane from a computation of the annulus diagram. in 6.5 the
calculation is generalized to finite temperature. Finally, in
section 6.6 we discuss the relation between s-branes and
D-instantons.

\subsec{The classical closed string field}

Consider the state \eqn\fdt{| C\rangle ={1 \over L_0+\bar
L_0}| B\rangle .} This can be viewed as a quantum state of a
single closed string. Alternately, since the states in the Fock
basis of the single closed string are identified as spacetime
components of the classical string field, $| C \rangle$ can be
viewed as a classical string field configuration. By construction
it obeys \eqn\fdtg{(L_0+\bar L_0)| C\rangle =| B\rangle .}
These are the linearized spacetime wave equations with sources for
the components of the classical string field. The source is the
boundary state $| B \rangle$ whose support is confined to the
brane. Hence we conclude that $| C \rangle$ is the linearized
classical closed string field sourced by the brane.

For ordinary static supersymmetric D-branes, it has been
explicitly verified that $| C \rangle$ as defined in \fdt\ reproduces the
linearized dilaton, metric and RR  fields sourced by the
brane \pdv. When the brane is static, the Fock component states in
$| B\rangle$ carry no $p^0$, and
all of its components are off-shell. A unique static $| C
\rangle$ may then be determined from \fdt. Time-dependent boundary
states differ crucially in this regard. They have components with
non-vanishing $p^0$ which correspond to on-shell closed string
states and hence are annihilated by $L_0+\bar L_0$. It follows
that $ L_0+\bar L_0$ is not invertible and  $| C \rangle$ is
not unambiguously determined from $| B \rangle $. This is just
the usual problem of specifying the homogenous part of the
solution emanating from a time-dependent source. Some additional
boundary conditions must be specified.\foot{We note that this
closed string ambiguity remains even after we have
fixed the quantum state of the open strings on the brane.}

For general $\lambda$, the Fock components of the string field
wave equation \fdtg\ are of the form \eqn\jjx{\eqalign{
\eta^{ab}\p_a\p_b \phi (\vec x, t)&=\delta^{25-p}(\vec x) {\cos^2
(\pi \lambda) \over 1+\sin^2( \pi \lambda)+2\sin(\pi \lambda)
\cosh t }\cr &=\delta^{25-p}(\vec x)\left({1 \over 1+ e^{t}\sin
\pi \lambda}+{1 \over 1+ e^{-t}\sin \pi \lambda}-1\right).}}  This
follows from \fz\ after replacing the operator $X^0$ by its
eigenvalue $t$.  Here $\vec x$ are the transverse spatial dimensions, and
longitudinal spatial dimensions are suppressed. For notational
simplicity we henceforth specialize to $p=22$, so that only four
spacetime dimensions are relevant.
For $\lambda=\half$ the wave equation \jjx\ reduces to
\eqn\jsjx{ \eta^{ab}\p_a\p_b \phi
(\vec x, t)=2\pi i \delta^3(\vec x)\sum_{m=-\infty}^\infty
\delta(t+\pi i +2m \pi i).} Since the source on the right hand
side vanishes for real $t$, an obvious solution of \jsjx\ for real
$t$ is simply \eqn\phj{\phi (\vec x, t)=0.} This is the solution
implicit in \SenNU.
However, there is another solution which "knows" about
the sources at imaginary $t$.
Recall that
the wave equation with a delta function source \eqn\jdx{
\eta^{ab}\p_a\p_b \phi (\vec x, t)=\delta^3(\vec x) \delta(t-t_0)}
is solved by Feynman propagator \eqn\pyu{\phi(\vec x,t)=-\Delta_F(\vec
x, t;\vec 0, t_0)={i \over 4 \pi^2}\lim_{\epsilon \to 0}
{1\over(t-t_0)^2-r^2 +i\epsilon },} where $r^2=\vec x^2$.
Continuing this to imaginary $t_0$ we find that \jjx\ is solved by
\eqn\xyt{\eqalign{\phi(\vec x,t)&=-2\pi i\sum_{m=-\infty}^\infty
\Delta_F(\vec x, t;\vec 0, \pi i +2\pi m i )\cr &={1 \over 2 \pi}
\sum_{m=-\infty}^\infty {1 \over r^2 -(t-\pi i -2\pi m i )^2}.}}
Since the denominator is
non-vanishing for real $(r,t)$, we have set $\epsilon$ to zero
here.
Performing the sum over $m$ yields \eqn\wki{\phi(\vec
x,t)=-{1\over 4 \pi r}\bigl( \tanh {r+t \over 2}+\tanh {r-t \over
2}\bigr).} Note that at large $t$ and fixed $r$ this vanishes
exponentially. On the other hand, at fixed time $t$ this has the
${1\over r}$ falloff at large $r$ characteristic of a static
source, yet it is nonsingular for all real $r,t$. With 25-p
transverse dimensions we would find a characteristic $1 \over
r^{23-p}$ falloff. Solutions of this general form were discussed
in \mgas.

In section 6.3 we will compute the annulus diagram
connecting an sD-brane and an ordinary D-brane using a
straightforward adaptation of the standard string theory
prescription \PolchinskiRQ,
which involves a Euclidean continuation of the
worldsheet field $X^0$. Since this calculation gives a definite
answer, it must contain an implicit prescription for inverting
$L_0+\bar L_0$.  We shall find that at large separation the
graviton falls off like $1 \over r^{23-p}$. This indicates that
non-trivial solutions of the form \wki\ rather than the trivial
solution \phj\ are implicit in this formulation of worldsheet
string theory.

We note that at $\lambda=\half$, the branes have no support at
real $t$, and therefore there can be no on-shell open strings 
propagating at real time. Hence the problem \StromingerPC\ of a
Hagedorn-like divergence in open string pair production
disappears.

\subsec{RR field strength and s-charge}
In this subsection we will use \fdt\ to determine the RR field sourced
by an sD-brane in superstring theory. We will find that it
carries one unit of ``s-charge''  defined below as an  integral
of the RR field strength over a complete spacelike surface.

In the presence of an s-brane described by the boundary deformation
\supers, the source for RR fields is proportional to \SenFTTM
\eqn\rrsource{ \sin(\pi\lambda) \left[ {e^{X^0/\sqrt{2}}\over
1+\sin^2(\pi\lambda) e^{\sqrt{2}X^0}} - {e^{-X^0/\sqrt{2}}\over
1+\sin^2(\pi\lambda) e^{-\sqrt{2}X^0}} \right] }
up to an overall normalization, which
can be determined as follows. If we analytically
continue
\eqn\asd{
\lambda\to -i\lambda,~~~ X^0\to X^0+\pi i/\sqrt{2},
}
the boundary interaction \supers\ becomes $\lambda\psi^0 \cosh
X^0/\sqrt{2}$, which describes the tachyon rolling over the
barrier. The corresponding s-brane should carry $\pm 1$ unit of
RR charge, with the sign depending on the sign of $\lambda$. The
integral of the source is now $ {\sqrt{2}\pi}\, {\rm
sign}(\lambda)$, which determines the normalization factor to be
$(\sqrt{2}\pi)^{-1}$ times the unit RR charge.

Let us now return to the case \supers. After Euclidean continuation $X^0\to
i\phi$, the RR source can be written as
\eqn\asd{
{1\over\sqrt{2}\pi}\sin(\pi\lambda)\left[\sum_{n=0}^\infty (-1)^n
\sin^{2n}(\pi\lambda) e^{i(2n+1)\phi/\sqrt{2}} - \sum_{n=0}^\infty
(-1)^n \sin^{2n}(\pi\lambda) e^{-i(2n+1)\phi/\sqrt{2}} \right]
.}
When $\lambda=1/2$, this is the source corresponding to an array
of branes and anti-branes located along the Euclidean time axis
\eqn\asd{-i\sqrt{2}
\sum_{n=-\infty}^\infty (-1)^n \delta(\sqrt{2}\phi+\pi +2n\pi)
}
The wave equation is then \eqn\cwaveeqn{
\partial^a\partial_a C_{9-p,\cdots,9}= \sqrt{2}\sum_{n=-\infty}^\infty (-1)^n \delta(\sqrt{2}X^0+i\pi
+2\pi ni)\delta(\vec{r}_{\perp}) .}
For simplicity, we will consider the case of an
s5-brane in type IIB theory, so that there are 4 transverse directions.
As in the previous section, we can solve the wave equation in Euclidean
space and analytically continue back to find \eqn\cfield{\eqalign{
C_{4\cdots9} &= {i\over 2\pi^2} \sum_{n=-\infty}^\infty
{(-1)^n\over (\sqrt{2}t+\pi i+2\pi n i)^2-2r^2} \cr &= {1\over
8\sqrt{2}\pi^2 r} \left[ {1\over \cosh{r-t\over \sqrt{2}}} -
{1\over \cosh{r+t\over \sqrt{2}}} \right] .}} The RR potential for
sp-branes, $p\ne 5$, in type IIB theory can be obtained
from the above by, say, taking derivatives in $r$.

The s-brane with $\lambda=1/2$ has no RR source located at real
time, so the RR flux through any transverse spacelike 3-surface is
conserved.  The conserved charge is 
\eqn\conserved{ Q_s =
\int_{\Sigma_3} *dC .} If we take $\Sigma_3$ to be the plane
located at $X^0=t$ and extending in $X^{1,2,3}$ directions, then
\eqn\dcfield{ (dC)_{04\cdots9}={\partial\over\partial_t}
C_{4\cdots9} = {1\over 16\pi^2 r} \left[ {\sinh{r-t\over
\sqrt{2}}\over \cosh^2{r-t\over \sqrt{2}}} + {\sinh{r+t\over
\sqrt{2}}\over \cosh^2{r+t\over \sqrt{2}}} \right] } yields
\eqn\conhalf{ \eqalign{ Q_s&=\int_0^\infty dr 4\pi r^2
(dC)_{04\cdots9} ={1\over 2}~. } } We conclude that the
$\lambda={1\over2}$ closed string configuration carries half a
unit of spacelike Ramond-Ramond charge.

\subsec{Long-range graviton/dilaton fields} In this subsection we compute
the long-distance Coulomb fields for the graviton and dilaton
sourced by the sp-brane.  We will restrict to the bosonic case,
although the generalization to the superstring is straightforward.

The spatial components of the boundary state of an sp-brane are
\eqn\gda{ |B\rangle_{\vec{X}} = {T_{p+1}\over 2}
\delta^{24-p}(\vec{x}_\perp) \exp\left( -\sum_{n=1}^\infty S_{ij}
a_{-n}^i \tilde{a}_{-n}^i \right) |0\rangle} where $T_{p+1}$ is
the tension of the D$(p+1)$-brane and $S_{ij}~(1\leq i,j\leq 25)$
is given by \eqn\gdb{ S_{ij} = (\delta_{\alpha\beta},
-\delta_{ab}) } where $\alpha,\beta$ ($a,b$ ) label the directions
with Neumann (Dirichlet) boundary conditions.

 The relevant parts
of the time component of the
 boundary state are \SenNU\
\eqn\gdstate{ |B\rangle_{X^0} = f(X^0) |0\rangle + a_{-1}^0
\tilde{a}_{-1}^0 g(X^0) |0\rangle + \cdots } where \eqn\gdfg{
\eqalign{ f(X^0) &= {1\over 1+ e^{X^0}\sin \pi\lambda} + {1\over
1+e^{-X^0}\sin \pi\lambda} -1, \cr g(X^0) &= 1+\cos(2\pi\lambda) -
f(X^0),} }and $a_{-1}^0$ is an oscillator in the expansion of
$X^0$. Combining \gda\ and \gdstate, we get the total source for
graviton and dilaton \eqn\gdsource{ |B\rangle={T_{p+1}\over 2}
\delta^{24-p}(\vec{x}_\perp) \left[ -S_{ij} a_{-1}^i
\tilde{a}_{-1}^j f(X^0) + a_{-1}^0 \tilde{a}_{-1}^0g(X^0) \right]
|0\rangle + \cdots } The massless part of the closed string field
is then \eqn\gdcfield{ |C\rangle = {T_{p+1}\over 2}V_{p+1}\int dt'
\Delta(\vec{X},X^0;0,t') \left[ -S_{ij} a_{-1}^i \tilde{a}_{-1}^j
f(t') + a_{-1}^0 \tilde{a}_{-1}^0g(t') \right] |0\rangle + \cdots
} where $V_{p+1}$ is the spatial volume of the s-brane. Our
prescription for the $t'$ integral employs the Euclidean imaginary
time axis. Defining \eqn\gdjmn{ J^{\mu\nu}(k) = \langle 0;k|
a_1^\mu \tilde{a}_1^\nu |C\rangle } one finds \eqn\gdj{\eqalign{
J_{ij}(x) &= -{T_{p+1}\over 2}V_{p+1}S_{ij} \int dt'
\Delta(\vec{x},t;0,t') f(t'),~~~\cr J_{00}(x) &= {T_{p+1}\over
2}V_{p+1}\int dt' \Delta(\vec{x},t;0,t') g(t'). }} 

For simplicity let us first consider the case $p=21$, where there
are 4 transverse directions to the s$p$-brane. For
$r=|\vec{x}|>|t|$,\foot{The full solution is discontinuous (but obeys the 
wave equation) on the light cone for $\lambda<\half$. The 
$\lambda=0$ case corresponds to a D-brane with some radiation inside the
light cone.}\eqn\gdjval{ \eqalign{ J_{ij}(\vec{x},t) &=
-{T_{p+1}V_{p+1}\over 2}{S_{ij} \over 4\pi r} \left[ {1\over
1+e^{t-r}\sin \pi\lambda} + {1\over 1+e^{-t-r}\sin\pi\lambda}-1
\right], \cr J_{00}(\vec{x},t) & = - {T_{p+1}V_{p+1}\over 2}
{1\over 4\pi r} \left[ {1\over 1+e^{t-r}\sin \pi\lambda} + {1\over
1+e^{-t-r}\sin\pi\lambda}-2-\cos(2\pi\lambda) \right]. } } In the
limit of large $r$, we have \eqn\gdlong{J_{ij}\to
-{T_{p+1}V_{p+1}\over 2}{S_{ij}\over 4\pi r}, ~~~~ J_{00}\to
{T_{p+1}V_{p+1}\over 2} {\cos(2\pi\lambda)\over 4\pi r }} For
general s$p$-branes \eqn\gdglong{ \eqalign{ &J_{ij}\to
-{N}_p{S_{ij}\over r^{22-p}}, ~~~~ J_{00} \to
{N}_p{\cos(2\pi\lambda)\over r^{22-p}}, \cr & {N}_p =
{T_{p+1}V_{p+1}\over 4} {\pi^{p-24\over 2}\Gamma\left({24-p\over
2}\right) }. } }
We see that in the string frame the graviton and dilaton fields 
fall off like ${1\over r^{22-p}}$. This is consistent with the results 
of the next subsection in which the force between an s-brane and a 
D-brane is computed.

\subsec{The annulus diagram}
 In this subsection we compute the
bosonic annulus diagram in the presence of an ordinary
D(p+1)-brane and sp-brane with general $\lambda$. $\lambda=0$
corresponds to two D(p+1)-branes while $\lambda=\half$ is an
sDp-brane and a D(p+1)-brane. We will deduce the long range force
from this computation and find that it is in agreement with the results
of the previous section.

Let us denote the boundary state associated to the D(p+1)-brane by
$|D_{p+1}\rangle$ and the boundary state associated to the
sp-brane with coupling $\lambda$  by $|s_p,\lambda \rangle$. They are
factorized as the product of
time components and spatial components as \eqn\asd{\eqalign{
& |D_{p+1}\rangle = |N\rangle^0\otimes |N\rangle^{1,\cdots,p+1}
\otimes |D\rangle^{p+2,\cdots,25}\otimes |{\rm ghost}\rangle, \cr
&|s_p\rangle = |B,\lambda \rangle \otimes |N\rangle^{1,\cdots,p+1}\otimes
|D\rangle^{p+2,\cdots,25}\otimes |{\rm ghost}\rangle. } }
Here $|B,\lambda \rangle$ is the zero-temperature boundary state (as in
\CallanUB, \SenNU\ and equation \cklmb)
\eqn\ssx{\eqalign{|B,\lambda \rangle &=P_\infty
e^{2\pi i \lambda J_1}|N\rangle,\cr &= 2^{-1/4} \sum_{j,m}
D^j_{m,-m}(2\pi \lambda) |j,m,m\rangle\rangle. }} Here
$J_1=\cosh X^0$, $P_\infty$ projects onto $p_L=p_R$ and
$D^j_{m,m'}$ is the $SU(2)$ representation matrix element
\eqn\xok{D^j_{m,m'}(2\pi \lambda) = \langle j,m|e^{i2\pi \lambda
J_1}|j,m'\rangle.} For the state $|j,m,m'\rangle\rangle$ the left
and right momenta are related by $p_L=2m', p_R=2m$.

The annulus diagram connecting  $|D_{p+1}\rangle$ and
$|s_p,\lambda \rangle$ is given by\foot{To compute the force
between the D-brane and the s-branes we must multiply this
expression by a factor of 2, since the string can stretch in
either orientation.} \eqn\rrc{Z_A=\int_0^\infty dt~ I_\lambda
(t),} with \eqn\amp{ I_\lambda (t)\equiv \langle s_p|
e^{-t(L_0+\widetilde{L}_0)} |D_{p+1}\rangle.} The integrand of
$Z_A$ is a product $I_\lambda(t)=I^0_\lambda(t)I^S(t)$ of
contributions from time components and spatial plus ghost
components. The contribution from these latter components is the
same as in the free theory, namely\foot{In general when there is
on-shell closed strings exchange one must include an ordering and
$i\epsilon$ prescription for the $t$ and $\vec p$ integrations.
However when one of the boundary states is a D-brane, local energy
conservation prohibits emission/absorption of an on-shell closed string.
} \eqn\sp{\eqalign{ I^S(t)&={1\over 2t} V_{p+1} \int {d^{p+1}{\vec
p}\over (2\pi)^{p+1}}e^{-{2\pi {\vec p}^2\over t}- {y^2\over 2\pi
t} } \sum e^{-2\pi(h_i-1)/t} \cr &={V_{p+1} \over 2t}
\left({8\pi^2\over t}\right)^{-(p+1)/2} e^{ -{y^2\over 2\pi t} }
\eta(i/t)^{-23} \cr &= {V_{p+1}\over 2} (8\pi^2)^{-(p+1)/2}
t^{(p-24)/2}e^{-{y^2\over 2\pi t}} \eta(it)^{-23}. }} The timelike
components were computed in \CallanUB\ (up to the factor of $i$
from analytic continuation of the volume) as \eqn\jhg{
I^0_\lambda(t)=\langle B|e^{-\pi t(L_0+\widetilde{L_0})}|N\rangle=
{iV_0\over\sqrt{2}} \sum_{j=0,1,\cdots} D_{00}^j(2\pi \lambda)~
\chi_j^{Vir}(e^{-2\pi t}),} where \eqn\wsz{ D_{00}^j(2\pi \lambda)
= {1\over j!} {d^j\over d\xi^j} \left[\xi^j(1-\xi)^j\right],~~~
\xi = \sin^2(\pi \lambda),} and $V_0$ is the real volume in the
$X^0$ direction and \eqn\cftp{ \chi_j^{Vir}(q) =
q^{-1/24}(q^{j^2}-q^{(j+1)^2}) \prod_{n=1}^\infty {1\over 1-q^n}}
is the Virasoro character.  The total integrand  from large $t$
goes as \eqn\rdzx{ I_\lambda(t)=i{V_0 V_{p+1}\over 2} 
{(8\pi^2)^{-(p+1)/2}}
\int dt~ t^{(p-24)/2}
e^{-y^2/2\pi\alpha't} (e^{2\pi t}+ 23 +\cos(2\pi
\lambda)+\cdots).} When $\lambda=0$ we have two D(p+1)-branes and
\rdzx reduces to the usual expression. As usual the force between
the D-brane and s-brane is obtained by differentiating with
respect to $y$. As $\lambda$ ranges from 0 to $\half$ and the
D-brane goes to an s-brane and then an sD-brane, the force
decreases by a factor of ${11 \over 12}$.

\subsec{The finite-temperature annulus } In this subsection we
compute the bosonic annulus diagram at general $\lambda$ and
temperature $T={1 \over 2\pi n}$. We will find that the $T\to 0$
limit reproduces the results of the previous section.

At finite temperature $T={1 \over 2\pi n}$ one naturally computes
the Euclidean thermal partition function. This is obtained by Wick
rotation  $X^0\to i\phi$ with the Euclidean time $\phi$
compactified on a circle of radius
$n$\refs{\plctmp,\PolchinskiRQ}. We consider a D(p+1)-brane
located at $X^m=0$, $m=p+2,\cdots, 25$, with world-volume
extending in the $X^1,\cdots,X^{p+1}$ directions and wrapped
around the $\phi$ circle. There is also an  sp-branes located at
$X^m=y^m$ and  parallel to the D-brane in the spatial directions.
When $\lambda$ vanishes, we simply have the Euclidean annulus
connecting two finite temperature D(p+1)-branes. The open string
1-loop calculation gives (before integrating over $t$)
\eqn\dbrane{\eqalign{ {\rm Tr}\, e^{-2\pi L_0/t} &= {1\over
\eta(i/t)} \sum_m e^{-{2\pi \over t} {m^2\over n^2} } \cr &=
{1\over \eta(i/t)} \vartheta(0,{2i\over n^2 t }) \cr &= {n\over
\sqrt{2}\eta(it)} \vartheta(0,in^2t/2). }} We will recover this
result below in the special case $\lambda=0$.

We will now turn to the Euclidean s-brane boundary state. At the
self-dual temperature ($n=1$), the time component of the boundary
state describing an s-brane is\CallanUB
\eqn\asd{
|B\rangle_{SU(2)} = e^{i2\pi \lambda J_1}|N\rangle_{SU(2)}. } For
other values of $n$ the boundary state describing an array of $n$
s-branes is, up to a normalization factor, simply the projection
of $|B\rangle_{SU(2)}$ onto allowed momentum and winding modes.
For $\lambda=\pm 1/2$, this boundary state describes $n$ Dirichlet
branes on a circle. The boundary state $|B\rangle$ at temperature
$T=1/2\pi n$ is \eqn\sbstate{\eqalign{ |B\rangle_{R=2\pi n}&=
2^{-1/4}\sum_{j=0,1/2,\cdots} P_n e^{i\theta^a J^a}
|j,m,-m\rangle\rangle \cr &= 2^{-1/4} \sum_{j} \sum_{m,w}
D^j_{m-wn,-m}(2\pi \lambda) |j,m,m-wn\rangle\rangle. }}
 At
$\lambda=\pm 1/2$ equation \sbstate\ simplifies to \eqn\iis{
2^{-1/4} \sum_{j=0,1/2,\cdots} e^{\pm i\pi j}
|j,m,m\rangle\rangle.} The Neumann boundary state at $T=1/2\pi n$
is \eqn\dxl{ |N\rangle_{R=2\pi n} = 2^{-1/4}
\sum_{j=0,1/2,\cdots}\sum_w |j,wn/2,-wn/2\rangle\rangle.} Here the
second sum is over allowed values of $w$, i.e. $wn$ is restricted
to be even (odd) when $j$ is integer (half integer). The time
component of the annulus amplitude is \eqn\time{\langle B|e^{-\pi
t(L_0+\widetilde{L_0})}|N\rangle_{R=2\pi n} = {n\over\sqrt{2}}
\sum_{j=0,1/2,1,\cdots} \sum_{w} D^j_{{wn\over 2},{wn\over
2}}(2\pi \lambda)~ \chi_j^{Vir}(e^{-2\pi t}).} The prefactor  $n$
comes from the volume of the $\phi$ zero mode.

The trivial case $\lambda=0$ corresponds to a Neumann boundary
condition. We should be able to recover from \time\ the answer for
ordinary D-branes \dbrane. We can write \time\ as \eqn\eqsum{
\langle N|e^{-\pi t(L_0+\widetilde{L_0})}|N\rangle_{R=2\pi n} =
{n\over\sqrt{2}} \sum_{j=0,1/2,1,\cdots} C_{n,j}\,
\chi_j^{Vir}(e^{-2\pi t}),} where \eqn\rkh{ C_{n,j} =
\sum_{wn/2\in\{-j,-j+1,\cdots,j\}}1.} Suppose $n$ is even. Then
the $wn/2$ are integers, so the contribution to \rkh\ comes from
integer values of $j$. Equation \eqsum\ can then be rewritten as
\eqn\asd{ {n\over \sqrt{2}\eta(it)} \sum_{j=0}^\infty
(C_{n,j}-C_{n,j-1}) e^{-2\pi j^2t} .} For $n$ even, \eqn\asd{
C_{n,j}-C_{n,j-1}=\left\{ \eqalign{&2, ~~ j=kn/2,k\geq 1; \cr &1,
~~ j=0; \cr &0,~~{\rm otherwise}} \right. } so we can evaluate the
sum \eqsum \eqn\asd{ {n\over
\sqrt{2}\eta(it)}\sum_{k=-\infty}^\infty e^{-\pi n^2k^2t/2} =
{n\over \sqrt{2}\eta(it)}\vartheta(0,in^2t/2) ,} which agrees with
\dbrane. A similar analysis yields the same expression for odd
$n$.

At the special values $\lambda=\pm 1/2$ only the $w=0$ sector will
contribute to the amplitude \time, so $j$ is therefore restricted
to be an integer. The amplitude \time\ then reduces to \eqn\spec{
{n\over\sqrt{2}} \eta(it)^{-1} \sum_{j=-\infty}^\infty e^{-2\pi t
j^2 + \pi i j} = {n\over\sqrt{2}\eta(it)}\vartheta(1/2,2it).}
Combining this with \sp, we find that the annulus partition
function at generic $\lambda$ and $n$ is \eqn\annu{\eqalign{
\int_0^\infty dt~ \langle s_p|& e^{-\pi t(L_0+\widetilde{L}_0)}
|D_{p+1}\rangle \cr &= \int_0^\infty dt {nV_{p+1}\over 2\sqrt{2}}
(8\pi^2)^{-{p+1\over 2}} t^{p-24\over 2} e^{-{y^2\over 2\pi t}}
\eta(it)^{-24} \cr & ~~~~ \sum_{j=0,1/2,1,\cdots} \sum_w
D^j_{{wn\over 2}, {wn\over 2}}(2\pi \lambda)~ (e^{-2\pi t
j^2}-e^{-2\pi t (j+1)^2}). }}

For $\lambda=\pm 1/2$ and all $n$, the contribution from large $t$
goes like \eqn\asymp{ nV_{p+1}\int dt~ t^{(p-24)/2}
e^{-y^2/2\pi\alpha't} (e^{2\pi t}+22+\cdots).} As usual, the first
term in the sum corresponds to the amplitude of tachyon exchange.
The graviton exchange amplitude, given by the second term in the
sum, falls off like $|y|^{p-22}=|y|^{2-(25-(p+1))}$, as in the
zero-temperature case. This agrees with \rdzx\ up to the factor of
$i$, which is due to the fact that \asymp\ comes from the
Euclidean rather than Lorentzian one loop diagram.

\subsec{S-branes and D-instantons}

In this subsection we discuss the relation between
s-branes and D-instantons.

Instantons fall into two general categories: those with and those without
a time reversal symmetry.  An example of the latter is the
Yang-Mills instanton. It describes tunneling between topologically distinct
vacua. Any attempt to continue it to real time yields an imaginary
solution. There is a topological Lorentzian configuration -- the creation and decay of a sphaleron
-- which interpolates between the distinct vacua, but it is not obtained by
analytic continuation of the Euclidean instanton solution.

The situation is different when there is a time reversal symmetry, as in the
Euclidean bubble describing the decay of the false vacuum. The analytic
continuation to real time yields a real solution describing a
contracting/expanding bubble of the true vacuum inside the false
vacuum. Indeed, the semi-classical decay process is quantum tunneling
followed by the real time expanding bubble solution. In this case both
the Euclidean and Lorentzian solutions are real and meaningful.

The sD-branes (and most of the s-branes) discussed herein are of
this latter character. They are processes in which the tachyon
rolls up one side of the barrier and back down the same side, and
so are time symmetric. While we have been focusing on the
Lorentzian solutions, there are also Euclidean solutions which
represent tunneling through the tachyon barrier at ${\cal T}=0$.
For example, we could consider a semi-classical process in which
the tachyon is incident on the barrier from ${\cal T}=-\infty$,
tunnels through it, and then proceeds to ${\cal T}=+\infty$. The
tunneling phase of this evolution would be described by the
Euclidean continuation of the s-brane. It connects two
time-symmetric solutions in which ${\cal T}$ bounces off the
barrier.  Alternately the periodically identified instanton can be
interpreted as a finite-temperature tunneling in which the energy
comes from the thermal bath. Such tunneling processes could 
become important near the Hagedorn transition. 

There are also of course non-time symmetric solutions with large enough
energy to classically pass over the barrier. These solutions, 
largely the focus of \mgas, will result in a
change in the RR
s-charge $Q_s$ evaluated at $t=\pm \infty$. 
They are analogs of baryon-number-violating
sphaleron creation/decay in
the standard model. They do not have a real continuation to
Euclidean space (at least by the usual method). 
In the superstring, one has a $\cosh$ rather than
a $\sinh$ in \supers, while in the bosonic string one has a $\sinh$ rather
than a $\cosh$ in \tis.

\newsec{Timelike Holography}

 An interesting potential application of s-branes is to the
 problem of finding a string theory configuration with a timelike
 holographic dual. We close this paper with some speculation on
 this topic.

 The sD-brane boundary state
 for the superstring at temperature $T={1 \over 2\sqrt{2}\pi n}$
 describes $2n$ Euclidean D-branes spaced along a circle at
 intervals $\sqrt{2}\pi$. According to \cwaveeqn, these have
 alternating RR charge, and hence are really $n$ D-branes and $n$
 anti-D-branes. One may also consider beginning with a boundary interaction on $N$
 coincident Lorentzian D(p+1)-branes, in which case the individual Euclidean (anti) Dp-branes
become replaced by a collection of $N$ coincident (anti)
Dp-branes.

According to the discussion of Section 4, this finite temperature
configuration determines an effective long-distance  Euclidean
field theory. This field theory evidently contains $2n$
supersymmetric $U(N)$ gauge theories. Additionally, the
$\sqrt{2}\pi$ spacing is precisely such that the would-be
tachyonic open string connecting a D-brane and an anti-D-brane is
massless. (A similar massless mode appears in the bosonic
sD-brane.) This couples the $U(N)$ theories with bifundamentals in
a manner that breaks supersymmetry.

Assuming they do exist, what could the dual supergravity solutions
be? In part these should be determined by the symmetries.  A
number of potentially dual solutions have appeared in the
literature. Many of them have an R-symmetry corresponding to
Lorentz transformations transverse to the brane. This symmetry is
clearly spontaneously broken in all the s-branes discussed herein,
and hence should not appear in the supergravity solution. Some
solutions that do have the appropriate symmetries have appeared in
\refs{\LuER, \QuevedoTM,\BurgessGG}. Particularly intriguing in
this connection is a class of solutions \BurgessGG\ that exhibit
thermal particle production and are periodic in Euclidean time.

 \centerline{\bf Acknowledgements} We are grateful to
M. Fabinger, S. Gubser, M. Gutperle, J. Karczmarek, M. Kruczenski,
F. Leblond, J. Maldacena,
S. Minwalla, L. Motl, R. Myers, A. Sen, T. Takayanagi
and J. Wacker for useful conversations. This work was
supported in part by DOE grant DE-FG02-91ER40654.

\listrefs

\end
\subsec{Eliminating the Hagedorn divergence}
 It is interesting to note that at $\lambda=\half$, the problem of
Hagedorn type divergences in open string pair production disappear,
simply because there is no
brane present at real $(x,t)$ and therefore no open string pair
production.  This can also be seen from looking at the closed string
channel of the annulus.
The breakdown  of  string
perturbation theory at $T_H={ 1 \over 4\pi}$ can be seen \PolchinskiRQ\
from looking at the annulus diagram \eqn\taq{Z(T)=-\ln \langle
B_{2\pi/T}| \int dq e^{q(L_0+\bar L_0)} | B_{2\pi/T}
\rangle.} When $T=T_H$, a \fxz\  acquires a component with a zero
eigenvalue of $(L_0+\bar L_0)$. This leads to the divergence in
\taq\ which signals the breakdown of perturbation theory.

However for the special value $\lambda =\half$, this
divergence is absent. In that case $\cos \lambda \pi $ and \fxz\
vanish, so there are no massless modes in the closed string
channel of \taq.